\documentclass[twocolumn]{aastex631}

\usepackage{amsmath}
\usepackage{wrapfig}
\usepackage{pythonhighlight}

\usepackage{CJK}

\begin{document}
\begin{CJK*}{UTF8}{gbsn}

\title{Two Types of $1/f$ Range in Solar Wind Turbulence}

\author[0000-0001-9570-5975]{Zesen Huang (黄泽森)}
\affiliation{Department of Earth, Planetary, and Space Sciences, University of California, Los Angeles, CA, USA}
\author[0000-0002-2381-3106]{Marco Velli}
\affiliation{Department of Earth, Planetary, and Space Sciences, University of California, Los Angeles, CA, USA}
\author[0000-0003-4177-3328]{B. D. G. Chandran}
\affiliation{Space Science Center and Department of Physics, University of New Hampshire, Durham, NH 03824, USA}
\author[0000-0002-2582-7085]{Chen Shi (时辰)}
\affiliation{Department of Earth, Planetary, and Space Sciences, University of California, Los Angeles, CA, USA}
\author[0009-0009-0162-2067]{Yuliang Ding (丁宇量)}
\affiliation{Department of Earth, Planetary, and Space Sciences, University of California, Los Angeles, CA, USA}
\author[0000-0002-6276-7771]{Lorenzo Matteini}
\affiliation{Imperial College London, South Kensington Campus, London SW7 2AZ, UK}
\author[0000-0003-2054-6011]{Kyung-Eun Choi}
\affiliation{Space Sciences Laboratory, University of California, Berkeley, CA 94720, USA}

\begin{abstract}
The $1/f$ noise is a ubiquitous phenomenon in natural systems. Since the advent of space exploration, the $1/f$ range has been consistently observed in \textit{in situ} solar wind measurements throughout the heliosphere, sparking decades of debate regarding its origin. Recent Parker Solar Probe (PSP) observations near the Alfv\'en surface have revealed a systematic absence of the $1/f$ range {above $10^{-4}$ Hz} in pristine solar wind, providing a unique opportunity to investigate its origin in solar wind turbulence. Despite numerous observations of the $1/f$ range at varying frequencies, no study has systematically examined its properties across different solar wind conditions. Here, we identify two distinct types of $1/f$ ranges in solar wind turbulence: the fast/Alfv\'enic wind type and the slow/mixed wind type. The fast/Alfv\'enic type appears to be an intrinsic feature of Alfv\'enic turbulence, while the slow/mixed type resembles classical flicker noise. For the fast/Alfv\'enic type, we find a near-perfect WKB evolution of the frequency-averaged fluctuation amplitude and an intriguing migration pattern in frequency space. For the slow/mixed type, we examine the solar cycle dependence of the $1/f$ noise using the OMNI-LRO dataset spanning solar cycles 22 to 25. We also analyze the autocorrelation function of the magnetic field vectors and identify a clear relationship between the $1/f$ range and the decline in correlation, as well as unexpected resonance peaks in the autocorrelation function.

\end{abstract}


\section{Introduction} \label{sec:intro}
$1/f$ noise (also known as pink noise or flicker noise) was first identified in the 1920s in electrical circuits \citep{johnson_schottky_1925,schottky_uber_1918,schottky_zur_1922}. Since then, it has been observed across a wide range of natural systems \citep{press_flicker_1978,milotti_1f_2002}, including vacuum tubes \citep{brophy_variance_1969}, undersea ocean currents \citep{taft_equatorial_1974}, and radio signals \citep{voss_1fnoise_1975}. In the context of solar wind turbulence, $1/f$ noise—often referred to as the $1/f$ range—typically appears in the low-frequency portion of the trace magnetic power spectral density (PSD). It has been observed in fast solar wind \citep{bavassano_statistical_1982,denskat_statistical_1982,wu_spectral_2025}, Alfv\'enic slow wind \citep{perrone_highly_2020,dorseth_1f_2024}, compressible non-Alfv\'enic slow wind \citep{bruno_low-frequency_2019}, and in long intervals of solar wind with mixed sources \citep{matthaeus_turbulent_1986,wang_1f_2024,pradata_observations_2025}. {Because the $1/f$ range is also present in the photosphere and corona at similar frequencies, \cite{wang_1f_2024} directly relate the solar wind $1/f$ range to the solar dynamo.} The characteristics of the $1/f$ range—such as its frequency and turbulence amplitude—vary depending on the type of solar wind \citep{damicis_alfvenic_2021,damicis_alfvenic_2025}. Nevertheless, its pervasive presence has made the $1/f$ range a fundamental feature of the solar wind turbulence PSD \citep{tu_mhd_1995,bruno_solar_2013}.

The $1/f$ range is also often referred to as ``energy-containing range'' or ``energy-injection range'' because of the $1/f$ scaling. This can be explained using the following formula:
\begin{eqnarray}\label{eq:P(f)}
    P(f) \ \mathrm{d}f = P(f) \cdot f \ \mathrm{d}\ln f
\end{eqnarray}
where $P(f)$ is the PSD. If $P(f)\propto 1/f$, the energy distributes evenly across the logarithmically spaced frequencies. Thus, the $1/f$ range has long been regarded as the energy reservior that supplies energy to facilitate the turbulence cascade in the inertial range, which eventually dissipates into internal energy via various mechanisms \citep{bowen_constraining_2020,shi_oblique_2020}.

The origin of $1/f$ range, however, remains an actively debated topic. For instance, \citet{matthaeus_low-frequency_1986} suggested that the $1/f$ range originates from log-normally distributed fluctuations on the solar surface \citep[see also][]{matthaeus_density_2007,wang_1f_2024,pradata_observations_2025}. \citet{velli_turbulent_1989} and \citet{verdini_origin_2012} proposed that it arises from the turbulent cascade of Alfv\'en waves in the expanding solar wind. Additionally, \citet{chandran_parametric_2018} argued that the $1/f$ range in the fast wind results from the parametric decay-induced inverse cascade of Alfv\'en waves. Other explanations include self-organized criticality \citep{bak_self-organized_1987}, inverse cascade of magnetic helicity \citep{galtier_introduction_2016}, phase mixing \citep{magyar_phase_2022}, saturation of fluctuations \citep{matteini_1_2018}, {and solar dynamo \citep{nakagawa_dynamics_1974,bourgoin_magnetohydrodynamics_2002,dmitruk_magnetic_2014}}. Thus, the origin of the $1/f$ range remains a compelling topic of research within the community.

The launch of Parker Solar Probe (PSP) \citep{fox_solar_2016,raouafi_parker_2023} in 2018 has provided an abundance of new data and revealed numerous surprising discoveries in the inner heliosphere. New observations from near the Alfv\'en surface, the boundary separating the solar corona and the solar wind \citep{kasper_parker_2021,chhiber_alfven_2024}, have revealed surprising features of the $1/f$ range: the $1/f$ range is systematically absent {above $10^{-4}$ Hz} in the most pristine solar wind, where the PSD instead exhibits a shallow-inertial double power law in the low-frequency regime \citep{huang_new_2023}. Tracking the evolution of a single coronal hole outflow, the low-frequency shallow spectrum gradually evolves into the $1/f$ range \citep{davis_evolution_2023}. As indicated by (\ref{eq:P(f)}), the shallow-inertial double power law suggests a concentration of fluctuation energy around 2-minute periods. This period was interpreted as the launch periods of the coronal hole Alfv\'en waves \citep{huang_dominance_2024}, which is consistent with the swaying motions of open coronal hole field lines observed by SDO-AIA \citep{morton_basal_2019}, and the coronal doppler velocities derived from DKIST Cryo-NIRSP measurements \citep{morton_origins_2025,morton_high-frequency_2025}.

The novel observations from PSP have rejuvenated this classical topic \citep{wu_spectral_2025,dorseth_1f_2024,pradata_observations_2025,wang_1f_2024,damicis_alfvenic_2025,shaikh_turbulence_2024,matteini_alfvenic_2024,morikawa_solar_2023,larosa_evolution_2024,terres_investigating_2024,wang_coronal_2024}. The newly discovered shallow-inertial double power law appears to support the dynamical formation of the $1/f$ range. However, as argued in \cite{wang_1f_2024}, dynamic formation theories encounter causality issues when applied to the $1/f$ range observed at ultra-low frequencies ($f < 10^{-5}$ Hz). This creates a real dilemma in identifying a one-size-fits-all mechanism for the origin of the $1/f$ range in solar wind turbulence. The solution may lie in the intrinsic nature of the $1/f$ range itself. Indeed, the $1/f$ range observed in the compressible, non-Alfv\'enic slow wind seems to differ substantially from its counterpart in the fast Alfv\'enic wind \citep{bruno_low-frequency_2019,damicis_alfvenic_2021}. Furthermore, the $1/f$ ranges reported in \cite{matthaeus_low-frequency_1986} and \cite{matthaeus_density_2007} were computed from long-interval measurements of the solar wind at 1 AU, which were mixtures of different types of solar wind. This suggests that the $1/f$ range observed in different types of solar wind may have fundamentally different origins. Nonetheless, to the best of our knowledge, no dedicated effort has been made to systematically analyze the different $1/f$ ranges across a broad spectrum of solar wind conditions.

In this study, we systematically identify and analyze the $1/f$ range at various locations in the heliosphere over the past four solar cycles. The paper is organized as follows. In the next section, we introduce the spacecraft data and analysis methods. Section~\ref{sec:result} presents a comprehensive overview of the $1/f$ range in solar wind turbulence across different heliospheric locations, examines the radial evolution of the $1/f$ amplitude in fast/Alfv\'enic wind, and investigates the temporal variation of the $1/f$ range at L1 over four solar cycles. We also explore the relation between the $1/f$ range and autocorrelation function of magnetic field vector. In Section~\ref{sec:discussion}, we discuss the constraints on the origin of $1/f$ range and broader implications of our findings. We conclude with a summary of the study.

\section{Data and Method}\label{sec:data}

In this study, we analyze data from PSP \citep{fox_solar_2016,bale_highly_2019,kasper_alfvenic_2019,kruparova_quasi-thermal_2023}, Solar Orbiter \citep{muller_solar_2020}, Helios \citep{porsche_helios_1981}, Ulysses \citep{balogh_magnetic_1992,wenzel_ulysses_1992,mccomas_three-dimensional_2003}, WIND \citep{harten_design_1995}, and OMNI (LRO) \citep{king_solar_2005}. In this study, we employ the analysis method used in \cite{huang_dominance_2024}. Given a magnetic field vector time series:
\begin{eqnarray}
    \vec B(t) = (B_x(t), B_y(t), B_z(t))
\end{eqnarray}
The Fast-Fourier Transformation (FFT) \citep{harris_array_2020} is applied, and we obtain $\tilde{B}_x(f)$, $\tilde{B}_y(f)$, $\tilde{B}_z(f)$ in frequency space. The trace magnetic power spectral density is defined as:
\begin{eqnarray}
    P(f) = |\tilde{B}_x(f)|^2 + |\tilde{B}_y(f)|^2 + |\tilde{B}_z(f)|^2
\end{eqnarray}
Due to the noisy nature of the FFT spectrum, it is further smoothed using a fixed-size sliding window in logarithmic frequency space. Furthermore, based on (\ref{eq:P(f)}), $P(f) \cdot f$ represents the distribution of fluctuation energy in logarithmic frequency space \citep{huang_dominance_2024}. Thus, we present $P(f)\cdot f$ to visualize the distribution of fluctuation energy. Notably, this method was first employed in \cite{matthaeus_low-frequency_1986} to facilitate the identification of $1/f$ range. For the majority of this study, $P(f)\cdot f$ is used as the primary representation of the trace magnetic PSD.

\begin{figure*}
    \centering
    \includegraphics[width=1.0\textwidth]{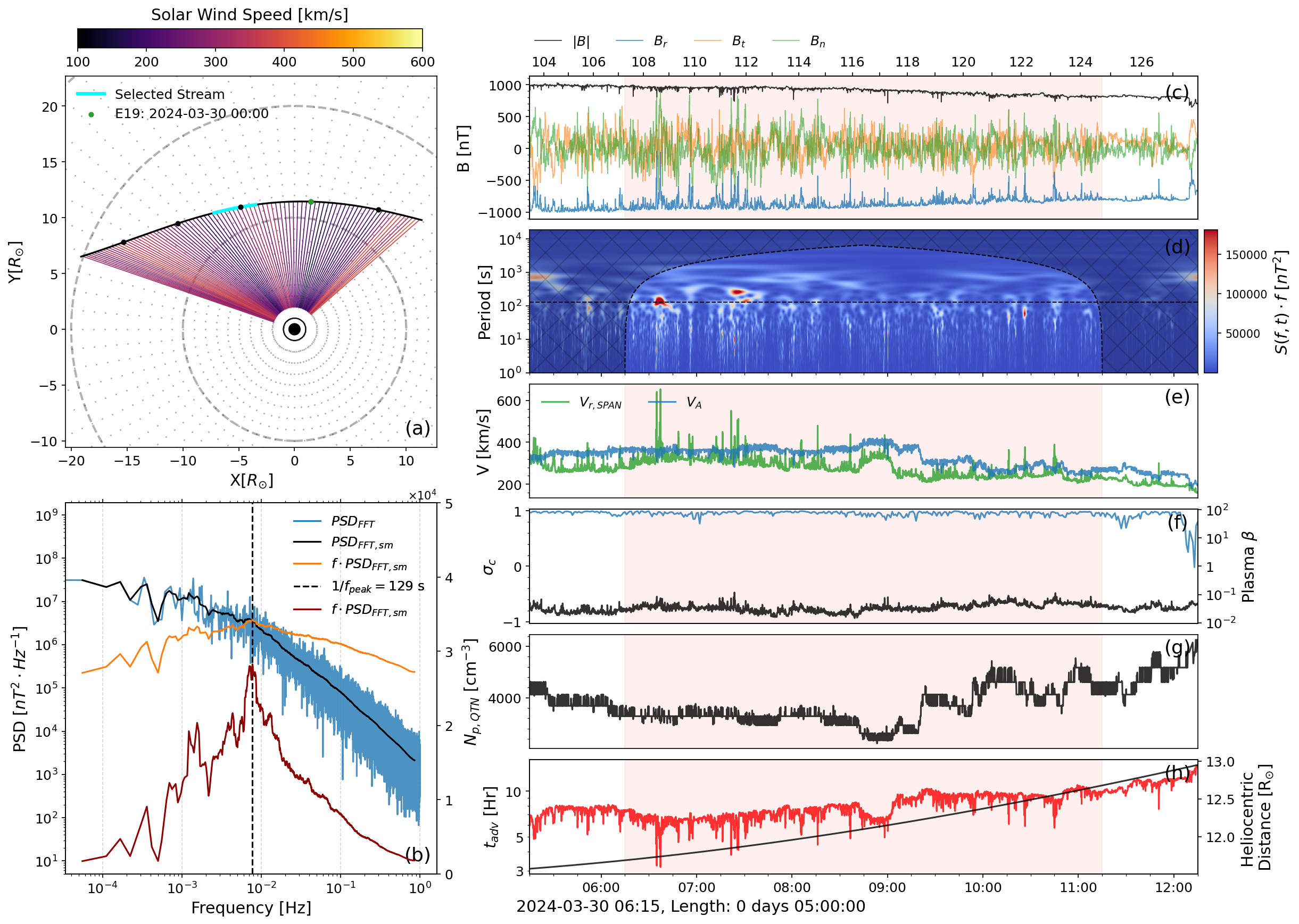}
    \caption{Example interval from PSP E19 perihelion. (a) Trajectory of PSP in Carrington corotating frame. Black dots are plotted every eight hours. Green dot indicates the entering direction of the spacecraft. The interval is highlighted with the cyan bar. The radial dotted lines are plotted every 5 degree longitude. The dashed circles are plotted every 10 $R_\odot$. (b) Trace magnetic PSD of the interval. (c) R-T-N components and magnitude of magnetic field. (d) Trace magnetic wavelet spectrogram. (e) Radial solar wind speed from SPAN-ion $V_{r,SPAN}$ and local Alfv\'en speed $V_A$ estimated from QTN. (f) 1-minute rolling averaged Cross helicity $\sigma_c$ (left) and proton plasma $\beta$ (right). (g) Electron number density from QTN. (h) Solar wind advection time $t_{adv}=(R-R_{\odot})/V_r$ (left) and heliocentric distance (right).}
    \label{fig:example}
\end{figure*}

Figure~\ref{fig:example} presents a representative spectrum from the perihelion of PSP E19. In panel (a), PSP enters from the right and rapidly traverses different solar wind streams. Panel (b) displays the trace magnetic PSD computed via FFT ($PSD_{FFT}$, blue), which is further smoothed to obtain $PSD_{FFT,sm}$. To illustrate the distribution of fluctuation energy, $PSD_{FFT,sm}$ is rectified by multiplying by $f$, shown in orange. For geometric interpretation, $PSD_{FFT}\cdot f$ is plotted again in a linear-log format using the right-hand-side y-axis (brown line). {Note that the orange line has been shifted in y-axis for illustration purpose, and the brown line is plotted with the original values.} According to (\ref{eq:P(f)}), the area under the dark-red curve represents the magnetic fluctuation energy within a given frequency range. Panels (c) through (h) confirm that the selected interval lies below the Alfv\'en critical surface and is dominated by large-amplitude, spherically polarized, outward-propagating Alfv\'en waves, {some of which are referred to as ``switchbacks'' in recent studies} \citep[see e.g.][]{bale_highly_2019,bale_interchange_2023,drake_switchbacks_2021,larosa_switchbacks_2021,dudok_de_wit_switchbacks_2020,zank_origin_2020,shi_patches_2022}. Furthermore, panel (d) implies that most of the fluctuation energy is concentrated in short-lived wave bursts with periods of approximately two minutes. For additional details on this interval, see \cite{huang_dominance_2024}.

To compare the behaviors of $1/f$ range with other statistical quantities, here we compute the autocorrelation of magnetic field vectors. Specifically, we calculate the averaged time-lagged correlation of magnetic field vectors:
\begin{eqnarray}\label{eq:autocorrelation}
    R(\mathrm{d}t)
    = 
    \left\langle
        \frac{\vec B(t)\cdot \vec B(t+\mathrm{d} t)}{|\vec B(t)||\vec B(t+\mathrm{d} t)|}
    \right\rangle_{t}
\end{eqnarray}
where for every $\mathrm{d}t$, we randomly select 20000 pairs of $\vec B(t)$ and $\vec B(t+\mathrm{d} t)$. {It should be noted that, unlike the common approach in turbulence analysis \citep{matthaeus_evaluation_1982,pope_turbulent_2015}, we do not subtract the mean magnetic field before computing the autocorrelation function. This choice is intentional, as retaining the mean field allows for a more direct interpretation of the correlation between magnetic field vectors over different time lags. Here, $R(\mathrm{d}t)$ represents the average value of $\cos\theta = \vec B_1 \cdot \vec B_2/(|\vec B_1| |\vec B_2|)$, where $\theta$ is the angle between two magnetic field vectors separated by a time lag $\mathrm{d}t$. Previous studies have shown that the distribution of $\cos\theta$ or $\theta$ provides valuable insights into the properties of solar wind turbulence \citep{matteini_rotation_2019,larosa_evolution_2024}. The results are presented in Figure~\ref{fig:correlation}.}

\section{Results}\label{sec:result}

\subsection{$1/f$ Range Throughout the Heliosphere}

\begin{figure*}
    \centering
    \includegraphics[width = 0.8\textwidth]{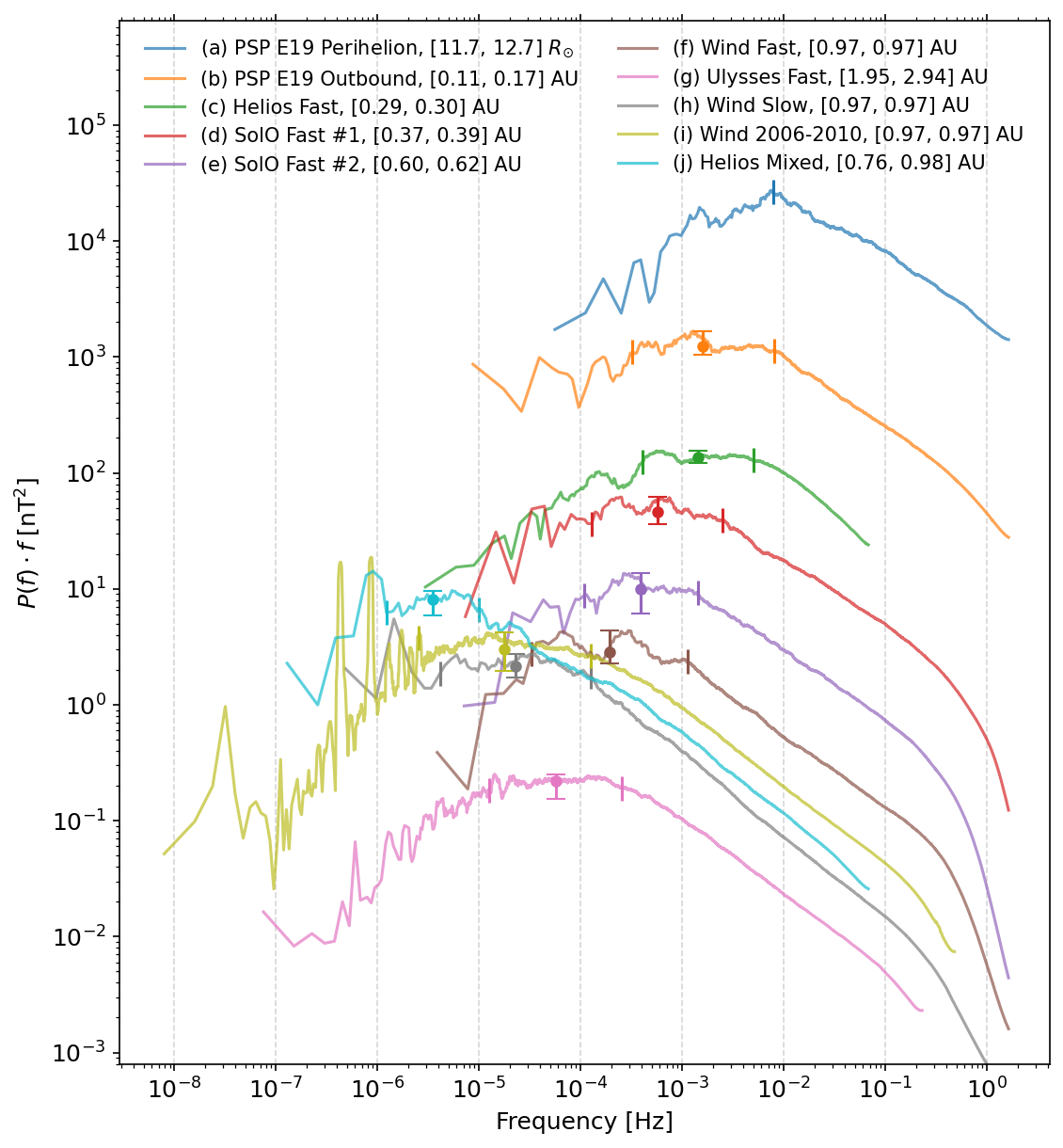}
    \caption{Overview of $1/f$ range in {rectified} trace magnetic PSD of solar wind turbulence. (a) Pristine Alfv\'enic (fast) wind from perihelion of PSP E19; (b) Alfv\'enic wind from $\pm$ 5 days of E19 perihelion; (c) Fast wind from outbound of E19; (d) Helios fast wind; (e) Solar Orbiter fast wind; (f) WIND fast wind; (g) Ulysses polar fast wind; (h) Helios mixed solar wind; (i) OMNI-LRO mixed solar wind; (j) WIND slow wind. {The vertical bars indicates the $1/f$ ranges, and the errorbars shows the averaged values ($\pm 1 \sigma$) of $P(f)\cdot f$ within the $1/f$ ranges.}}
    \label{fig:panorama}
\end{figure*}

In Figure~\ref{fig:panorama}, we present a panoramic view of the $1/f$ range in the trace magnetic PSD of solar wind turbulence, encompassing a broad range of heliospheric locations based on previous studies. \textbf{(a)} to \textbf{(g)} are PSDs compiled from various Alfv\'enic or fast wind streams observed by PSP, Helios, Solar Orbiter, WIND, and Ulysses. \textbf{(h)} to \textbf{(j)} represent slow wind or extended intervals mixing different solar-wind streams, collected near 1 AU by Helios and WIND. Each PSD serves as a characteristic example associated with specific solar-wind conditions. The detailed explanations for each line are as follows:

\begin{itemize}
    \item \textbf{(a)} Pristine Alfv\'enic fast wind measured at the perihelion of PSP E19 (2024-03-30/06:15 + 5 Hr, $\langle V_r\rangle = 279.64 \mathrm{km/s}$), identical to Figure~\ref{fig:example}. The $1/f$ range is absent, and instead the spectrum is shallower, indicating a concentration of fluctuation energy near the spectral break (around 2 minutes). {Note that in Figure~\ref{fig:example}, the $P(f)\cdot f$ (orange line) is shifted vertically for illustration purposes, so the y-axis values are not directly comparable.}
    \item \textbf{(b)} Pristine Alfv\'enic fast wind from outbound section of PSP E19 (2024-03-31/16:00 + 32 Hr, $\langle V_r\rangle = 418.75 \mathrm{km/s}$), dominated by Alfv\'enic fluctuations with cross-helicity close to unity. 
    \item \textbf{(c)} Helios-2 fast-wind PSD measured around 0.3 AU  (1976-04-13 to 04-23, $\langle V_r\rangle = 726.55 \mathrm{km/s}$, similar to Fig2-2(c) of \cite{tu_mhd_1995}). The $1/f$ range aligns closely with PSP observations made at smaller heliocentric distances.
    \item \textbf{(d)} Solar Orbiter fast wind measured at around 0.38 AU (2022-10-23/06:30 + 38 Hr, $\langle V_r\rangle = 511.02 \mathrm{km/s}$). The $1/f$ range starts to migrate to lower frequencies compared to Helios observations.
    \item \textbf{(e)} Solar Orbiter fast wind measured at around 0.61 AU (2024-03-05/09:00 + 39 Hr, $\langle V_r\rangle = 487.27 \mathrm{km/s}$). The $1/f$ range migrates to even lower frequencies.
    \item \textbf{(f)} WIND fast wind PSD at 1 AU  (2007-01-17 00:00 + 72 Hr, $\langle V_r\rangle = 646.83 \mathrm{km/s}$, identical to Fig 3(b) of \cite{bruno_low-frequency_2019}). The $1/f$ range migrates to significantly lower frequencies compared to PSP, but the range size maintains.
    \item \textbf{(g)} Ulysses polar fast wind PSD  (1994-06-01 to 1994-10-01, $\langle V_r\rangle = 769.67 \mathrm{km/s}$). The $1/f$ range has shift one whole decade lower compared to PSP.
    \item \textbf{(h)} WIND slow-wind PSD (2009-05-30/00:00 + 24 day, $\langle V_r\rangle = 315.65 \mathrm{km/s}$, identical to Fig 2(b) of \cite{bruno_low-frequency_2019}). A well-formed $1/f$ range is present, differing from the frequency range trend set by (a) to (g).
    \item \textbf{(i)} Four year worth of WIND measurements at L1 point (2006-01-01 to 2010-01-01, $\langle V_r\rangle = 279.64 \mathrm{km/s}$). Similar to (h), it features an extended $1/f$ range around $10^{-5}$ Hz, and displays additional imprints of periodicities from the Carrington rotation. This interval is similar to Figure 1(e) of \cite{matthaeus_low-frequency_1986}.
    \item \textbf{(j)} Helios mixed solar wind PSD compiled from half an orbit (1976-10-01 to 1977-06-01, $\langle V_r\rangle = 411.73 \mathrm{km/s}$). Despite the mixture of fast, slow, and solar wind transients, a prominent $1/f$ range persists around $10^{-5}$ Hz, but is located at lower frequencies and higher amplitude compared to (h) and (i).
\end{itemize}

Overall, the $1/f$ range can be categorized into two distinct types: (1) Fast/Alfv\'enic type; and (2) Slow/Mixed type. The first type {shows} a clear trend from (a) through (g), whilst the second type are located in much low frequencies as shown by (h) to (j). The results are generally consistent with previous studies \citep{bruno_solar_2013,dorseth_1f_2024,bruno_low-frequency_2019}, and reveal numerous intriguing new features. In this study, we focus on the distribution of fluctuation energy instead of the power spectral scaling index of the PSD. Thus, we present the spectra using $P(f)\cdot f$, which visualizes the distribution of fluctuation energy across the logarithmic frequencies. Moreover, $P(f)\cdot f$ has the same dimension of $\delta B^2$, which can be interpreted as the fluctuation amplitude for the given frequency. In addition, we select the longest possible intervals whilst ensuring the homogeneity of the physical properties of the solar wind streams. This enables us to capture the entirety of the energy containing range, and even include much of the {low frequency declining part} of the PSD, which is rarely reported in previous studies.

There are some notable features from type (1): The $1/f$ range is missing for pristine solar wind measured near the Alfv\'en surface \citep{huang_dominance_2024}. As the solar wind evolves with increasing heliocentric distance, the $1/f$ range rapidly forms, and occupies approximately 1.5 decade in logarithmic scale. {As the solar wind travels outwards, the $1/f$ range gradually migrate to lower and lower frequencies at a steady pace, and the Alfv\'en wave amplitude steadily drops. }


Type (2), on the other hand, seems to be located at much lower frequencies for similar heliocentric distances, e.g. comparing (f) and (h). From (i) we can find various resonance peaks from the Carrington rotation, and the $1/f$ range extends to the lowest expected frequencies \citep{matthaeus_low-frequency_1986,wang_1f_2024,pradata_observations_2025}. Interestingly, by mixing half an orbit of Helios measurement, $1/f$ range shows up at frequencies even lower than the ones found at L1 point. 


\subsection{Radial Evolution of $1/f$ Range in Fast/Alfv\'enic Wind}\label{subsec:radial alfven}

\begin{figure*}
    \centering
    \includegraphics[width = 1.0\textwidth]{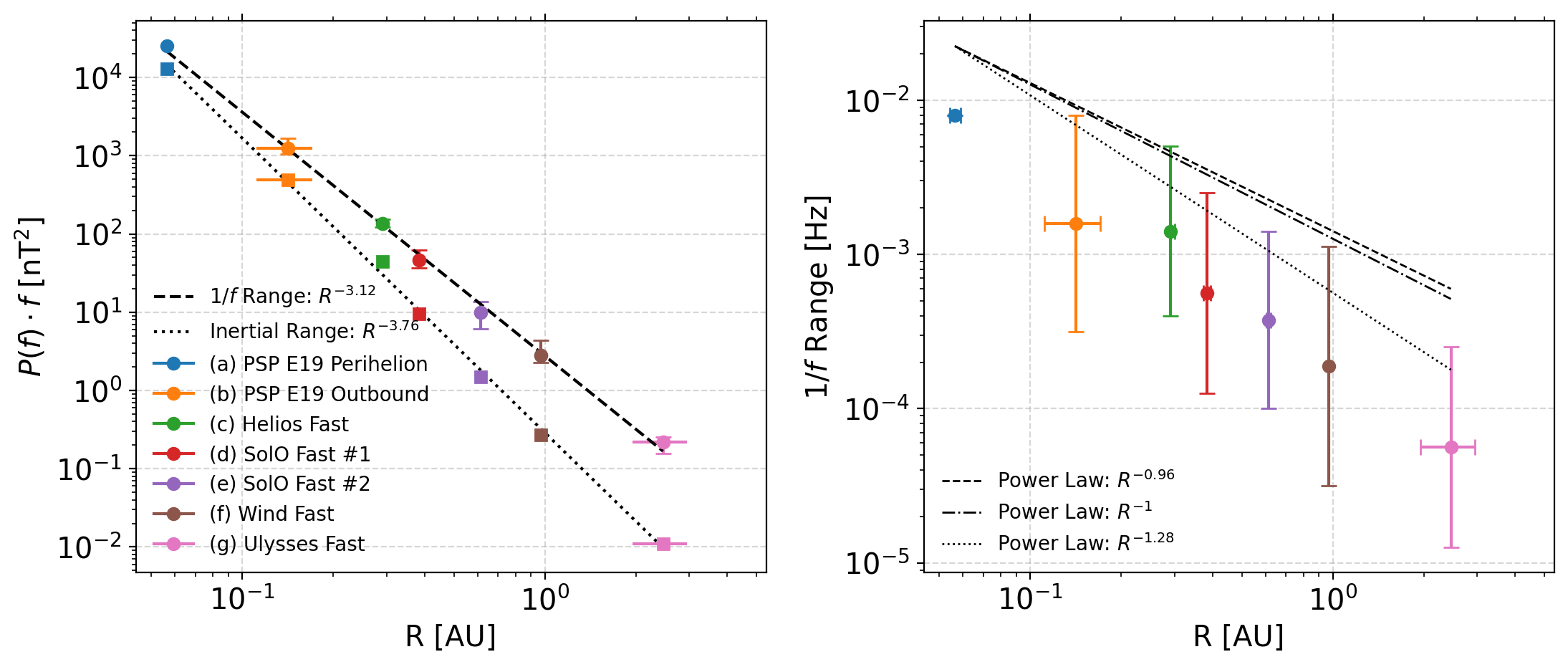}
    \caption{Left: Radial dependence of $P(f)\cdot f$ for (a) through (g). The circles are averaged values of $P(f)\cdot f$ within the $1/f$ range, highlighted with vertical bars in Figure~\ref{fig:panorama}. The squares are values of $P(f)\cdot f$ at $f=10^{-1.5}$ Hz. For all intervals, $10^{-1.5}$ Hz is located deep within the inertial range. Right: The radial evolution of the frequencies of the $1/f$ range. {The horizontal errorbars indicate the heliocentric distance range of the intervals.}}
    \label{fig:WKB}
\end{figure*}

From Figure~\ref{fig:panorama} we can see a clear trend of the $1/f$ range in Fast/Alfv\'enic wind, where it gradually migrate to lower frequencies whilst steadily decreases the fluctuation amplitude. Thus, to study the radial evolution of the PSD, we computed the averaged value of $P(f)\cdot f$ within the $1/f$ range highlighted in Figure~\ref{fig:panorama} (for (a), {the value is the peak of the spectrum at the vertical bar}) and at the fixed $f=10^{-1.5}$ Hz in the inertial range, and plot as a function of $R$. The results are shown in Figure~\ref{fig:WKB}. The averaged values are interpreted as the representative Alfv\'en wave amplitude. For (a), the value is taken from the maximum point of the spectrum, located around $10^{-2}$ Hz. The wave amplitudes follow a clear power law that is close the Wentzel-Kramers-Brillouin (WKB) prediction of  $\delta B^2 \propto R^{-3}$ {\citep[see e.g.][also see appendix]{whang_alfven_1973,belcher_alfvenic_1971,velli_propagation_1993,tenerani_evolution_2021,huang_dominance_2024}. }


As expected, the turbulence amplitudes in the inertial range possess a slightly steeper power law \citep{tu_mhd_1995}. Consider an inertial range scaling of $f^{-\alpha}$, the difference between the radial power law exponent serves as a good estimation of the migration rate of the $1/f$ range. Assuming that the inertial range drops like $R^{-a}$ and the $1/f$ range drops like $R^{-b}$, the gap between the two ranges will extend vertically at the rate of $R^{(b-a)}$. Thus, to concatenate the $f^{-1}$ range and $f^{-\alpha}$ range, the inertial range should extend to lower frequencies at the rate of $R^{-\frac{b-a}{\alpha - 1}}$. Using $a = 3.12, b=3.76$, the migration rate of the $1/f$ range is estimated to be $R^{-0.96}$ for $\alpha = 5/3$, and $R^{-1.28}$ for $\alpha=3/2$ \citep{kolmogorov_local_1941,kraichnan_inertialrange_1965,iroshnikov_turbulence_1964,goldreich_toward_1995,boldyrev_spectrum_2005,boldyrev_spectrum_2006,chandran_reflection-driven_2019}. Based on the inertial range scaling index in Table~\ref{tab:powerlawslopes}, the migration rate lies between these two values. The migration rate is visualized in the right panel of Figure~\ref{fig:WKB}, where the frequency ranges between the vertical bars in Figure~\ref{fig:panorama} are shown as errorbars. The low frequency spectral breaks are indicated with the top caps. This result is particularly interesting because the $1/f$ range starts to migrate to lower frequencies outside of {Alfv\'en surface}, and in the mean time it is very to close $R^{-1}$, which is indicative of geometric effect from solar wind expansion. Intriguingly, this result is sufficiently different from the low frequency spectral break moving rate estimated from Helios measurements \citep{bruno_solar_2013}, where it moves like $R^{-1.52}$. Note that the method used here is qualitatively similar to \cite{tu_power_1984} \citep[see also][]{tu_basic_1989,tu_model_1993,tu_mhd_1995}.

\begin{table}[ht]
\centering
\begin{tabular}{l c c}
\hline
Label  & Dataset        & Spectral Slope ($\alpha$) \\
\hline
(a)   & PSP E19 Perihelion    & -1.34 \\
(b)  & PSP E19 Outbound & -1.64 \\
(c)  &  Helios Fast& -1.80 \\
(d)  &  SolO Fast \#1     & -1.60 \\
(e)  &  SolO Fast \#2     & -1.63 \\
(f)  &  WIND Fast     & -1.62 \\
(g)  &  Ulysses Fast     & -1.64 \\
(h)  & WIND Slow& -1.67\\
(i)  & WIND 2006-2010& -1.65\\
(j)  & Helios Mixed & -1.83 \\
\hline
\end{tabular}
\caption{Spectral slopes ($\alpha$) for $P(f)$ near $10^{-1.5}$ Hz for different datasets.}
\label{tab:powerlawslopes}
\end{table}

\subsection{Solar Cycle Variation of $1/f$ Range at L1 Point}

\begin{figure*}
    \centering
    \includegraphics[width = 1.0\textwidth]{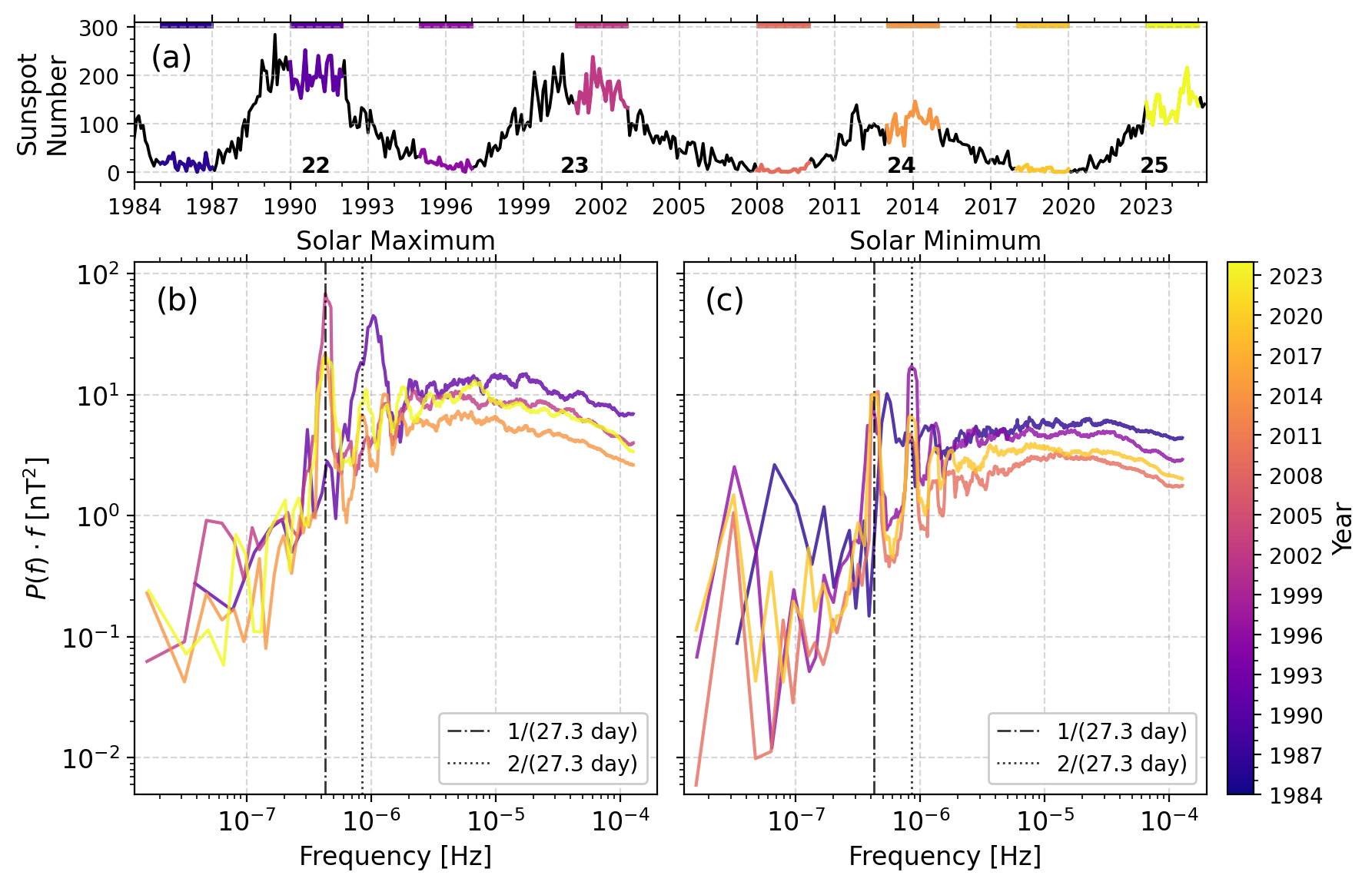}
    \caption{Time dependence of the OMNI (LRO) trace magnetic PSD over solar cycle 22, 23, 24 and 25. Each line represents two years worth of data. (a) Sunspot number from 1985-01 to 2025-03, where each of the selected time ranges are highlighted with the corresponding colors; (b) Solar Maximum; (c) Solar minimum. Dashed-dotted line: Carrington rotation. Dotted line: half-cycle Carrington rotation.}
    \label{fig:omni_seasonal}
\end{figure*}

The $1/f$ ranges found in slow/mixed solar wind, on the other hand, is much harder to study its radial evolution. This is because they are typically found at much lower frequencies compared to those in fast/Alfv\'enic wind, making it more difficult to collect a fair ensemble of intervals to make comparison. Nonetheless, the OMNI (LRO) dataset provides consistent and continuous maesurements of interplanetary magnetic field (IMF) at L1 point, which enables us to make temporal comparison of the $1/f$ range from mixed solar wind over the last four solar cycles (22 to 25). The results are shown in Figure~\ref{fig:omni_seasonal}.

We selected eight two-year long intervals, and each represents typical minimum or maximum phase of the each solar cycle. There are several notable features: (1) The spectral power is stronger during solar maxima compared to solar minima; (2) The Carrington rotation resonance peaks are more prominent during solar minimum than solar maximum, especially for the second harmonics; (3) The $1/f$ range is more well-formed during solar minima compared to those in solar maxima; (4) The $1/f$ range typically shows up in the frequencies higher than the Carrington rotation harmonics. The results shown here are consistent with previous observations using OMNI dataset and Messenger dataset \citep{matthaeus_low-frequency_1986,wang_1f_2024,pradata_observations_2025}.

\subsection{$1/f$ Range and Magnetic Field Autocorrelation}

\begin{figure*}
    \centering
    \includegraphics[width = 1.0\textwidth]{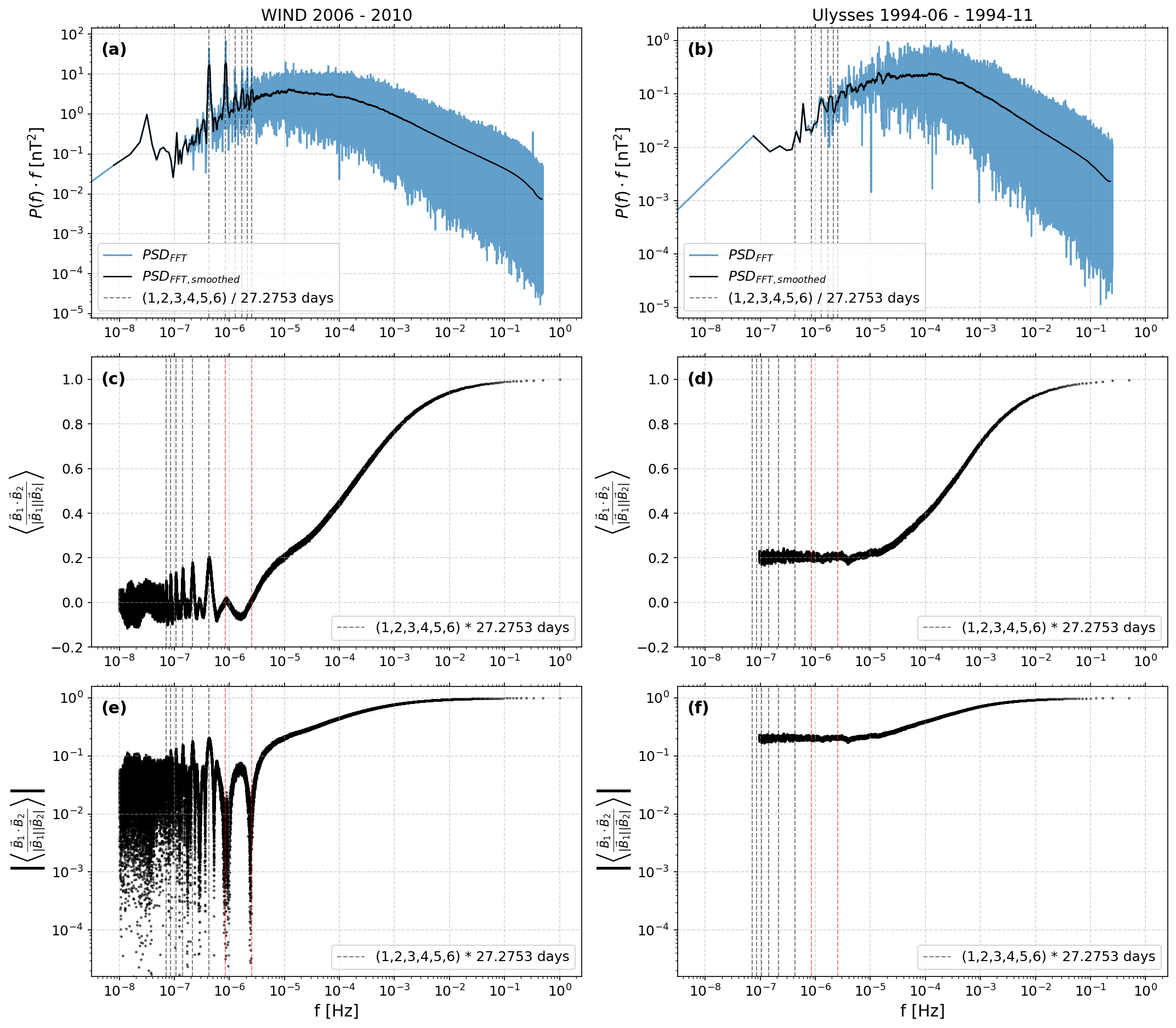}
    \caption{$1/f$ range and magnetic field vector correlation. \textbf{(a)} and \textbf{(b)} are identical to (f) and (g) in Figure~\ref{fig:panorama} respectively. The first six harmonics from Carrington rotation are highlighted with the dashed lines. \textbf{(c)} and \textbf{(d)} show the magnetic field vector autocorrelation as a function of $f=1/\mathrm{d}t$. Up to six inverses of Carrington rotations periods are highlighted with black dashed lines. The second and sixth harmonics are highlighted with the red dashed lines. \textbf{(e)} and \textbf{(f)} show the absolute value of the autocorrelation function.}
    \label{fig:correlation}
\end{figure*}

To investigate the origin of the $1/f$ range, it is instructive to examine its relation with other statistical quantities, among which the autocorrelation function of magnetic field vectors is particularly intriguing due to its interpretability. The formula is given in (\ref{eq:autocorrelation}). The autocorrelation function is computed using 20,000 randomly selected pairs of magnetic field vectors separated by each time lag $\mathrm{d}t$. The cosine values between each pair are averaged, and the absolute values are presented as a function of $f=1/\mathrm{d}t$. Note that the autocorrelation function used here differs from that in \cite{matthaeus_stationarity_1982}, where the mean magnetic field is subtracted. This distinction leads to substantially different behaviors and interpretations. The results are shown in Figure~\ref{fig:correlation}, where panels (f) and (g) from Figure~\ref{fig:panorama} are reproduced and overplotted with the original FFT spectrum. The four years of WIND data correspond to the solar minimum between solar cycles 23 and 24. The Ulysses data correspond to a single polar coronal hole outflow.

There are two main reasons for using the full magnetic field vectors: First, retaining the mean field simplifies the interpretation of the results. Second, the $1/f$ range, i.e. the energy-containing range, encompasses the majority of the fluctuation energy. The magnetic field fluctuations in the solar wind are typically large in amplitude, with the background field itself participating in the wave activity \citep[see e.g.][]{belcher_large-amplitude_1971,bale_highly_2019,matteini_1_2018}. Therefore, to identify potential connections to the $1/f$ range, it is more natural to retain the mean field in the autocorrelation calculation.

Several interesting features emerge from Figure~\ref{fig:correlation}. Notably, the autocorrelation function $R(t)$ starts at one at high frequencies and then starts to decrease. For both cases, the $1/f$ range corresponds to the frequency range where the autocorrelation function steadily declines. This indicates that the background magnetic field participates in the fluctuations within the $1/f$ range. As pointed out by \cite{matteini_1_2018}, the $1/f$ range corresponds to the saturation of magnetic field fluctuations, i.e., $\delta B/B \sim 1$, especially in fast/Alfv\'enic wind. The results here further confirm that as frequency decreases, the magnetic field vectors become increasingly randomized within the $1/f$ range, until reaching a randomization threshold (panel e) or saturation value (panel f). Interestingly, there's a ``bent'' in $R(t)$ curve near $10^{-4.5}$ Hz for WIND dataset (panel c), which happen to be in the middle of the $1/f$ range.

Additionally, in the WIND data PSD shown in Figure~\ref{fig:correlation}(a), there are six significant resonance peaks corresponding to the Carrington rotation. The first two peaks are particularly prominent, consistent with Figure~\ref{fig:omni_seasonal}, while the third through sixth harmonics are less pronounced. 
{Notably, a similar 13.5-day periodicity (the second resonance peak) has also been reported in previous study of the chromosphere, near-Earth solar wind, and interplanetary magnetic field observations \citep{mursula_135-day_1996}.} The presence of the third, fifth, and sixth peaks is unexpected, as during solar minimum the heliosphere is typically organized into two or four sectors \citep{hudson_solar_2014,choi_origin_2019}, which would not necessarily produce distinct resonance peaks at these harmonics. These resonance peaks are entirely absent in the Ulysses data, even after accounting for the differential rotation of the Sun. This absence is likely because the Ulysses measurements were taken from a single polar coronal hole outflow, where sector structures were absent.

The most intriguing features are the various resonant structures in $R(t)$ or $|R(t)|$. As shown in (c) and (e), there are numerous features at special frequencies. For example, up to six peaks related to the Carrington rotation can be identified, as indicated by the vertical dashed lines. Unlike the PSD, the resonances in $R(t)$ occur at integer multiples of the Carrington rotation period, leading to higher-order correlations. Interestingly, $R(t)$ approaches zero at certain time lags, indicating a balanced distribution of $\cos \theta = \vec B_1 \cdot \vec B_2/(|\vec B_1| |\vec B_2|)$, and implying that the magnetic field vectors become completely random at and only at these specific time lags. Moreover, there are two different types of zeros as indicated by the first and second red dashed lines. These are surprising findings, as they provides useful information about the randomness of the IMF over large time scales.

The first resonance coincides with 1/6 of a Carrington rotation, as indicated by the red dashed line. This likely implies that the heliosphere during solar minimum, on average, exhibits a three-sector structure, meaning that it has a roughly equal chance of exhibiting two-sector and four-sector configurations. Under these conditions, if one randomly samples pairs of IMF at L1 with a cadence of 1/6 of a Carrington rotation, the polarity is completely random due to the effective three-sector structure. The origin of other resonances appears to be more complex and is beyond the scope of this study.

In summary, the autocorrelation function provides some constraints on the origin of the $1/f$ range. For the $1/f$ range found in mixed-type solar wind, it is located below the first deep drop of $|R(t)|$, indicating that the $1/f$ range is confined to a single sector of solar wind. For the $1/f$ range observed by Ulysses, a similar property is seen: the $1/f$ range ends once $|R(t)|$ reaches a saturation value around 0.2, where the magnetic field vectors become completely random (aside from the uniform radial background field). This suggests that within the $1/f$ range, although $\delta B$ is saturated, the fluctuations remain correlated. Once the magnetic field loses this correlation, the fluctuation energy begins to decrease and the $1/f$ range terminates.

\section{Discussion}\label{sec:discussion}

\subsection{Migration of $1/f$ Range in the Fast Wind and Applicability of WKB Theory}

The WKB prediction of $\delta B^2 \propto R^{-3}$ is based on several assumptions: constant mass flux, Alfv\'en speed much less than the solar wind speed, and spherical expansion of the solar wind stream (see appendix). Most importantly, the WKB prediction is valid for a given frequency of the wave in the solar frame, which is determined at its source. However, in Figure~\ref{fig:panorama} and Figure~\ref{fig:WKB}, we computed the power law between $\delta B^2$ and $R$ using the averaged values estimated within the $1/f$ range. This approach is reasonable in the sense that $P(f)\cdot f$ is a good representation of the amplitude of the Alfv\'en wave due to the energy-containing nature of $1/f$ range. However, the corresponding frequencies for each of the intervals vary significantly as $R$ increases, violating the WKB assumptions. Nevertheless, $\delta B^2$ follows an almost perfect WKB prediction, and from (b) to (g), the size of the $1/f$ range seems irrelevant to its migration.

This leads to an obvious dilemma for the traditional interpretation of the migration of the $1/f$ range. The movement of the low-frequency spectral breaks to lower frequencies as $R$ increases has long been attributed to the turbulence cascade, which consumes energy in the $1/f$ range and gradually erodes it \citep{wu_energy_2021,wu_energy_2020,tu_mhd_1995,bruno_solar_2013}. One would expect such nonlinear erosion to shrink the size of the $1/f$ range and modify its radial decrease rate. However, the observations show that the $1/f$ range seems to migrate to lower frequencies as a whole, and decreases radially as dictated by linear geometric optics theory (WKB). Even more interestingly, Section \ref{subsec:radial alfven} points out that the migration starts outside of {Alfv\'en surface}, and at the same time the migration rate is almost {inverse} proportional to $R$ ($R^{-[0.95-1.26]}$, see section 3.2). This implies a geometric origin for the migration of the $1/f$ range due to solar wind expansion.

Nevertheless, a simple geometric picture for the migration rate faces difficulties. This stems from the interpretation of the frequencies in the PSD. Close to the Sun, the frequencies of the Alfv\'en waves—especially for (a) in Figure~\ref{fig:panorama}—can be interpreted as the launch frequency of the waves at the coronal base. {This is because PSP ($V_{r,PSP} \sim 100 \mathrm{km/s}$) is much slower than the radial group velocity of the Alfv\'en waves ($U+V_A\simeq 1000 \mathrm{km/s}$)}, and therefore the PSP frame can be considered quasi-static in the solar frame. The frequencies measured in the spacecraft frame can therefore be directly related to the launch frequency of the waves, modified by a negligible radial Doppler shift factor \citep[]{huang_dominance_2024}. Farther from the Sun, the scenario quickly becomes more complicated. For one thing, the Parker spiral starts to become non-negligible, and hence the Alfv\'en waves no longer propagate perfectly radially outward. In addition, the Alfv\'en speed becomes much smaller compared to the solar wind speed, meaning that the Alfv\'en waves are no longer effectively propagating, making them \textit{de facto} advective. Therefore, at 1 AU and beyond, it is more common to invoke the Taylor Hypothesis \citep{taylor_spectrum_1938,perez_applicability_2021} to interpret the temporal signals as passively convected spatial structures by the solar wind, thereby converting the frequency space into wavenumber space \citep[see e.g.][]{zank_spectral_2020,sioulas_magnetic_2022,sioulas_evolution_2023,mcintyre_properties_2023,zhao_turbulence_2024,zhu_radial_2025}. This could potentially be the direct cause for the migration of the $1/f$ range, because as the solar wind expands, the spatial structures naturally lengthen and hence the corresponding frequencies from solar wind convection decrease. However, the solar wind only expands in the perpendicular direction, and the solar wind only converts the radial wavenumber into frequency, where no expansion is expected. Nevertheless, the migration rate being close to {$R^{-1}$} outside of {Alfv\'en surface} makes it tempting to relate the observation to simple geometric effects. This opens an interesting topic for future studies.

\subsection{Two Types of $1/f$ Range}

Figure~\ref{fig:panorama} demonstrates that there are at least two types of $1/f$ range: (1) the fast/Alfv\'enic wind type and (2) the slow/mixed wind type. 

The fast/Alfv\'enic type seems to form dynamically near the Alfv\'en surface, as indicated by (a) and (b). And as the solar wind evolves, the $1/f$ range gradually migrates to lower frequencies and amplitudes drops per WKB prediction. Based on the current observational evidence, certain constraints can be put on the origin of either types of $1/f$ range. A few models have proposed the \textit{in situ} formation of the $1/f$ range: \cite{velli_turbulent_1989} and \cite{verdini_origin_2012} suggested that the reflection of Alfv\'en waves will lead to the formation of $1/f$ range; \cite{chandran_parametric_2018} proposed that the inverse cascade of Alfv\'en waves driven by parametric decay instability will create the $1/f$ range dynamically; \cite{magyar_phase_2022} showed that the $1/f$ range can be created by linear phase mixing process due to the inhomogeneity of solar corona. The major constraint is its formation location and formation rate. For example, observations suggests that the $1/f$ range in fast/Alfv\'enic wind typically forms near the Alfv\'en surface \citep{huang_new_2023,davis_evolution_2023,huang_dominance_2024}, where the concentrated Alfv\'en wave energy suddenly spread over at least a decade of frequency over the course of few hours in terms of advection time, e.g. from (a) to (b). This provide some constraints on the formation rate for the future simulation studies. Moreover, even though \cite{davis_evolution_2023} has tracked a single coronal hole outflow and showcased the dynamic formation process of $1/f$ range, more detailed and systematic observational studies on the \textit{in situ} formation $1/f$ range is needed to provide more rigorous constraints.

{The $1/f$ range in slow/mixed solar wind evidently has a different origin. Its low-frequency character requires much longer data intervals to adequately capture the $1/f$ range compared to the fast/Alfv\'enic type. Consequently, investigating its radial evolution is difficult, especially given the rapid motion of PSP. At present, a direct comparison of the radial evolution between the two types is not feasible.} It has been argued in \cite{wang_1f_2024} that causality issues arise when applying the dynamic formation mechanisms to the low frequency ($f \sim 10^{-5} Hz$) portion of the $1/f$ range. And it is true that a substantial part of $1/f$ range in this type resides below $10^{-5} Hz$. Thus, it is likely that the slow/mixed wind type of $1/f$ range is a remnant of solar surface structures {or solar dynamo} as suggested by various studies \citep{matthaeus_low-frequency_1986,matthaeus_density_2007,wang_1f_2024,pradata_observations_2025}. As shown in Figure~\ref{fig:omni_seasonal}, the $1/f$ range is more prominent during solar minimum than in solar maximum, suggesting that a more stable and consistent helmet streamer outflow favors the existence of $1/f$ range. Moreover, Figure~\ref{fig:correlation} indicates that the $1/f$ range corresponds to the frequency range where the magnetic field autocorrelation substantially drops. These evidence further justifies the name ``$1/f$ noise'' for the slow/mixed wind type of $1/f$ range because: (1) The signals within the $1/f$ range are not causally related; (2) The $1/f$ range is stronger when the signal is drawn from a stable stochastic source. The fast/alfv\'enic wind type of $1/f$ range, on the other hand, seems to be an intrinsic part of the Alfv\'enic turbulence, and thus is more appropriate for the name ``$1/f$ range''.


\section{Conclusions}

In this paper we discuss the two different types of $1/f$ range in the solar wind turbulence: (1) fast/{Alfv\'enic} type and (2) slow/mixed wind type. These two types of $1/f$ range exhibits significantly different properties, and hence they likely have different physical origins. The fast/Alfv\'enic wind type follows the WKB prediction but gradually migrates to low frequencies outside of {Alfv\'en surface} while maintaining its size in frequency space. It seems to be an intrinsic part of Alfv\'enic turbulence. The slow/mixed wind type shows strong variation over the solar activities but seems to be confined within one single solar wind sector. Thus it more likely to manifest the nature of uncorrelated noise. Both types of $1/f$ range exhibit a strong relation to the rapid declination of autocorrelation function of magnetic field vectors. Our results raise several intriguing questions for future studies:

\begin{enumerate}
    \item What are the origin of both types of $1/f$ range?
    \item What causes of the migration of fast-wind type $1/f$ range to lower frequencies?
    \item Is the $1/f$ range noise or part of solar wind turbulence? Is the fast wind type $1/f$ range an secondary inertial range of Alfv\'enic turbulence? How is it dynamically formed?
    \item How to explain the solar activity dependence of the $1/f$ range in slow/mixed wind?
\end{enumerate}

Parker Solar Probe has entered its final orbits with perihelia at 9.9 $R_{\odot}$, bringing the humanity beings one step closer to the source of the solar wind. Future observations from both PSP and Solar Orbiter will help us understand the deformation of the shallow-inertial double power law and the evolution of the $1/f$ range. Additionally, upcoming missions will help us constrain the origin of both types of $1/f$ range. For example, PUNCH can potentially help us investigate the origin of the low frequency $1/f$ noise from the fluctuations on the Sun \citep[see][ and references therein]{wang_1f_2024}. Europa Clipper can provide high- cadence measurements of the solar- wind magnetic field beyond 1AU during its journey to Jupiter \citep{pappalardo_science_2024}. Finally, {future missions such as the Solar Polar Orbiter \citep{national_academies_of_sciences_next_2025}} and the Chinese Solar Polar Orbiter \citep{deng_solar_2023} will likely provide imaging of the solar polar regions and \textit{in situ} measurements of polar solar wind at closer heliocentric distances compared to Ulysses \citep{bame_ulysses_1992}. Results from these missions will significantly advance our understanding of the origin of Alfv\'en waves and $1/f$ range in the solar wind.

\vspace{1em}
The code to calculate and smooth the PSD is available at: \href{https://pypi.org/project/SWTurbPy/}{https://pypi.org/project/SWTurbPy/} \citep{huang_swturbpy_2025}. {Hampel filter is used to remove obvious outliers in timeseries related instrument issues. It is available at:} \href{https://pypi.org/project/hampel-filter/}{https://pypi.org/project/hampel-filter/} \citep{huang_hampel_2025}. All spacecraft data are publicly available at CDAWeb \href{https://cdaweb.gsfc.nasa.gov/}{https://cdaweb.gsfc.nasa.gov/}.

\section*{Acknowledgements}
The authors acknowledge the following open-source packages: \cite{lam_numba_2015,harris_array_2020,hunter_matplotlib_2007,angelopoulos_space_2019,virtanen_scipy_2020}, and pycwt. This research was funded in part by the FIELDS experiment on the Parker Solar Probe spacecraft, designed and developed under NASA contract UCB \#00010350/NASA NNN06AA01C, and the NASA Parker Solar Probe Observatory Scientist grant NASA NNX15AF34G, and NASA HTMS 80NSSC20K1275. B.C. acknowledges support from NASA grant 80NSSC24K0171. C.S. acknowledges supported from NSF SHINE \#2229566 and NASA ECIP \#80NSSC23K1064. KEC was supported by NASA contracts 80NSSC22K0433, 80NSSC21K1770, and NASA’s Living with a Star (LWS) program (contract 80NSSC20K0218). We acknowledge UCLA OpenAI project for providing access to ChatGPT.

\clearpage
\appendix
\section{Conservation of Alfv\'en Wave Action in Expanding Solar Wind}\label{sec:transmission}
Consider an expanding flux tube formed by an open coronal hole outflow, and all quantities are function of heliocentric distance $R$. The cross-sectional area of the flux tube is $A$, and assume that the background magnetic field $B_0$ points radially outward. The energy density of outward($+$) and inward ($-$) propagating Alfv\'en wave are given by the following formula: 
\begin{eqnarray}
    e_{w}^\pm = \frac{1}{2}\rho \left(\frac{z^\pm}{2}\right)^2
\end{eqnarray}
where $\rho$ is the plasma density, and $z^{\pm}$ are Els\"asser variables, defined as $z^\pm = \delta \vec V \mp \delta \vec B/\sqrt{\mu_0 \rho}$. The wave action density is the adiabatic invariant of the Alfv\'en waves, which are defined as:
\begin{eqnarray}
    s^{\pm} = \frac{e_{w}^{\pm}}{\omega^{\pm}} = \frac{\frac{1}{2}\rho \left(\frac{z^\pm}{2}\right)^2}{\omega_0 \frac{V_A}{V_A \pm U}}
\end{eqnarray}
where $\omega_0$ is the launch frequency of the wave at coronal base, $U$ is radial speed of the solar wind, and $V_A=B_0/\sqrt{\mu_0 \rho}$ is the Alfv\'en speed. Note that $\omega^{\pm} = \omega_0 \frac{V_A}{V_A \pm U}$ is the doppler shifted wave frequency in the solar wind frame. The total wave action flux for an given cross section of the flux tube is:
\begin{eqnarray}
    S^{\pm}=s^{\pm} \cdot A \cdot (U\pm V_A) = \frac{1}{2} \rho \frac{(z^\pm/2)^2}{\omega_0\frac{V_A}{U\pm V_A}} \cdot (U\pm V_A) \cdot A
\end{eqnarray}
where $U\pm V_A$ are the group velocities of the outward and inward propagating Alfv\'en waves. It has been shown by \cite{velli_propagation_1993} that the net outward wave action flux is a conserved quantity in the MHD equations:
\begin{eqnarray}
    S^{+} - S^{-} = S_{\infty} = \mathrm{const}
\end{eqnarray}
where $S_{\infty}$ is the net outward wave action flux at infinity. Taking the Wentzel-Kramers-Brillouin (WKB) limit, we assume that the reflection of the outward propagating Alfv\'en wave is negligible, i.e. $S^{-} = 0$. In this case, $S^+$ is fully conserved:
\begin{eqnarray}\label{eq:outward_alfven}
    S^{+}=\frac{1}{2} \rho \frac{(z^+/2)^2}{\omega_0\frac{V_A}{U+ V_A}} \cdot (U+ V_A) \cdot A=\mathrm{const}
\end{eqnarray}
To reorganize the formula, we make one further assumption that the mass flux is conserved:
\begin{eqnarray}
    \rho U A = \mathrm{const}
\end{eqnarray}
and replace $(z^+/2)^2 = \delta B^2/(\mu_0\rho)$ because for outward propagating Alfv\'en waves $\delta \vec V = -\delta \vec B/\sqrt{\mu_0 \rho}$. (\ref{eq:outward_alfven}) can be reduced to:
\begin{eqnarray}
    \frac{\delta B^2}{\omega_0 \mu_0 \rho} \frac{(U+V_A)^2}{U V_A} = \mathrm{const}
\end{eqnarray}
For solar wind {far from the Alfv\'en surface}, we can further assume that $U\gg V_A$, $\rho \propto R^{-2}$, and $V_A=B_0/\sqrt{\mu_0 \rho}\propto R^{-1}$. Therefore, we have the WKB prediction of:
\begin{eqnarray}
    \delta B^2 \propto R^{-3}
\end{eqnarray}
The above derivation is a simplication of \citep{velli_propagation_1993} \citep[see also][and references therein]{belcher_alfvenic_1971,whang_alfven_1973,hollweg_alfven_1978,heinemann_non-wkb_1980,velli_waves_1991,huang_dominance_2024}.

\clearpage
\bibliography{sample631}{}

\begin{thebibliography}{}
\expandafter\ifx\csname natexlab\endcsname\relax\def\natexlab#1{#1}\fi
\providecommand{\url}[1]{\href{#1}{#1}}
\providecommand{\dodoi}[1]{doi:~\href{http://doi.org/#1}{\nolinkurl{#1}}}
\providecommand{\doeprint}[1]{\href{http://ascl.net/#1}{\nolinkurl{http://ascl.net/#1}}}
\providecommand{\doarXiv}[1]{\href{https://arxiv.org/abs/#1}{\nolinkurl{https://arxiv.org/abs/#1}}}

\bibitem[{Angelopoulos {et~al.}(2019)Angelopoulos, Cruce, Drozdov, Grimes, Hatzigeorgiu, King, Larson, Lewis, McTiernan, Roberts, Russell, Hori, Kasahara, Kumamoto, Matsuoka, Miyashita, Miyoshi, Shinohara, Teramoto, Faden, Halford, McCarthy, Millan, Sample, Smith, Woodger, Masson, Narock, Asamura, Chang, Chiang, Kazama, Keika, Matsuda, Segawa, Seki, Shoji, Tam, Umemura, Wang, Wang, Redmon, Rodriguez, Singer, Vandegriff, Abe, Nose, Shinbori, Tanaka, UeNo, Andersson, Dunn, Fowler, Halekas, Hara, Harada, Lee, Lillis, Mitchell, Argall, Bromund, Burch, Cohen, Galloy, Giles, Jaynes, Le~Contel, Oka, Phan, Walsh, Westlake, Wilder, Bale, Livi, Pulupa, Whittlesey, DeWolfe, Harter, Lucas, Auster, Bonnell, Cully, Donovan, Ergun, Frey, Jackel, Keiling, Korth, McFadden, Nishimura, Plaschke, Robert, Turner, Weygand, Candey, Johnson, Kovalick, Liu, McGuire, Breneman, Kersten, \& Schroeder}]{angelopoulos_space_2019}
Angelopoulos, V., Cruce, P., Drozdov, A., {et~al.} 2019, Space Science Reviews, 215, 9, \dodoi{10.1007/s11214-018-0576-4}

\bibitem[{Bak {et~al.}(1987)Bak, Tang, \& Wiesenfeld}]{bak_self-organized_1987}
Bak, P., Tang, C., \& Wiesenfeld, K. 1987, Physical Review Letters, 59, 381, \dodoi{10.1103/PhysRevLett.59.381}

\bibitem[{Bale {et~al.}(2019)Bale, Badman, Bonnell, Bowen, Burgess, Case, Cattell, Chandran, Chaston, Chen, Drake, de~Wit, Eastwood, Ergun, Farrell, Fong, Goetz, Goldstein, Goodrich, Harvey, Horbury, Howes, Kasper, Kellogg, Klimchuk, Korreck, Krasnoselskikh, Krucker, Laker, Larson, MacDowall, Maksimovic, Malaspina, Martinez-Oliveros, McComas, Meyer-Vernet, Moncuquet, Mozer, Phan, Pulupa, Raouafi, Salem, Stansby, Stevens, Szabo, Velli, Woolley, \& Wygant}]{bale_highly_2019}
Bale, S.~D., Badman, S.~T., Bonnell, J.~W., {et~al.} 2019, Nature, 1, \dodoi{10.1038/s41586-019-1818-7}

\bibitem[{Bale {et~al.}(2023)Bale, Drake, McManus, Desai, Badman, Larson, Swisdak, Horbury, Raouafi, Phan, Velli, McComas, Cohen, Mitchell, Panasenco, \& Kasper}]{bale_interchange_2023}
Bale, S.~D., Drake, J.~F., McManus, M.~D., {et~al.} 2023, Nature, 618, 252, \dodoi{10.1038/s41586-023-05955-3}

\bibitem[{Balogh {et~al.}(1992)Balogh, Beek, Forsyth, Hedgecock, Marquedant, Smith, Southwood, \& Tsurutani}]{balogh_magnetic_1992}
Balogh, A., Beek, T.~J., Forsyth, R.~J., {et~al.} 1992, Astronomy and Astrophysics Supplement Series (ISSN 0365-0138), vol. 92, no. 2, Jan. 1992, p. 221-236. Research supported by SERC., 92, 221

\bibitem[{Bame {et~al.}(1992)Bame, McComas, Barraclough, Phillips, Sofaly, Chavez, Goldstein, \& Sakurai}]{bame_ulysses_1992}
Bame, S.~J., McComas, D.~J., Barraclough, B.~L., {et~al.} 1992, Astronomy and Astrophysics Supplement Series (ISSN 0365-0138), vol. 92, no. 2, Jan. 1992, p. 237-265. Research supported by DOE., 92, 237

\bibitem[{Bavassano {et~al.}(1982)Bavassano, Dobrowolny, Fanfoni, Mariani, \& Ness}]{bavassano_statistical_1982}
Bavassano, B., Dobrowolny, M., Fanfoni, G., Mariani, F., \& Ness, N.~F. 1982, Solar Physics, 78, 373, \dodoi{10.1007/BF00151617}

\bibitem[{Belcher(1971)}]{belcher_alfvenic_1971}
Belcher, J.~W. 1971, Astrophysical Journal, vol. 168, p.509, \dodoi{10.1086/151105}

\bibitem[{Belcher \& Davis(1971)}]{belcher_large-amplitude_1971}
Belcher, J.~W., \& Davis, L. 1971, Journal of Geophysical Research, 76, 3534, \dodoi{10.1029/JA076i016p03534}

\bibitem[{Boldyrev(2005)}]{boldyrev_spectrum_2005}
Boldyrev, S. 2005, The Astrophysical Journal, 626, L37, \dodoi{10.1086/431649}

\bibitem[{Boldyrev(2006)}]{boldyrev_spectrum_2006}
---. 2006, Physical Review Letters, 96, 115002, \dodoi{10.1103/PhysRevLett.96.115002}

\bibitem[{Bourgoin {et~al.}(2002)Bourgoin, Mari{\'e}, P{\'e}tr{\'e}lis, Gasquet, Guigon, Luciani, Moulin, Namer, Burguete, Chiffaudel, Daviaud, Fauve, Odier, \& Pinton}]{bourgoin_magnetohydrodynamics_2002}
Bourgoin, M., Mari{\'e}, L., P{\'e}tr{\'e}lis, F., {et~al.} 2002, Physics of Fluids, 14, 3046, \dodoi{10.1063/1.1497376}

\bibitem[{Bowen {et~al.}(2020)Bowen, Mallet, Bale, Bonnell, Case, Chandran, Chasapis, Chen, Duan, Dudok~de Wit, Goetz, Halekas, Harvey, Kasper, Korreck, Larson, Livi, MacDowall, Malaspina, McManus, Pulupa, Stevens, \& Whittlesey}]{bowen_constraining_2020}
Bowen, T.~A., Mallet, A., Bale, S.~D., {et~al.} 2020, Phys. Rev. Lett., 125, 025102, \dodoi{10.1103/PhysRevLett.125.025102}

\bibitem[{Brophy(1969)}]{brophy_variance_1969}
Brophy, J.~J. 1969, Journal of Applied Physics, 40, 3551, \dodoi{10.1063/1.1658236}

\bibitem[{Bruno \& Carbone(2013)}]{bruno_solar_2013}
Bruno, R., \& Carbone, V. 2013, Living Reviews in Solar Physics, 10, 2, \dodoi{10.12942/lrsp-2013-2}

\bibitem[{Bruno {et~al.}(2019)Bruno, Telloni, Sorriso-Valvo, Marino, Marco, \& D{\textquoteright}Amicis}]{bruno_low-frequency_2019}
Bruno, R., Telloni, D., Sorriso-Valvo, L., {et~al.} 2019, Astronomy \& Astrophysics, 627, A96, \dodoi{10.1051/0004-6361/201935841}

\bibitem[{Chandran(2018)}]{chandran_parametric_2018}
Chandran, B. D.~G. 2018, Journal of Plasma Physics, 84, \dodoi{10.1017/S0022377818000016}

\bibitem[{Chandran \& Perez(2019)}]{chandran_reflection-driven_2019}
Chandran, B. D.~G., \& Perez, J.~C. 2019, Journal of Plasma Physics, 85, 905850409, \dodoi{10.1017/S0022377819000540}

\bibitem[{Chhiber {et~al.}(2024)Chhiber, Pecora, Usmanov, Matthaeus, Goldstein, Roy, Wang, Thepthong, \& Ruffolo}]{chhiber_alfven_2024}
Chhiber, R., Pecora, F., Usmanov, A.~V., {et~al.} 2024, Monthly Notices of the Royal Astronomical Society: Letters, 533, L70, \dodoi{10.1093/mnrasl/slae051}

\bibitem[{Choi \& Lee(2019)}]{choi_origin_2019}
Choi, K.-E., \& Lee, D.-Y. 2019, Solar Physics, 294, 44, \dodoi{10.1007/s11207-019-1433-7}

\bibitem[{D{\textquoteright}Amicis {et~al.}(2021)D{\textquoteright}Amicis, Perrone, Bruno, \& Velli}]{damicis_alfvenic_2021}
D{\textquoteright}Amicis, R., Perrone, D., Bruno, R., \& Velli, M. 2021, Journal of Geophysical Research: Space Physics, 126, e2020JA028996, \dodoi{10.1029/2020JA028996}

\bibitem[{D{\textquoteright}Amicis {et~al.}(2025)D{\textquoteright}Amicis, Velli, Panasenco, Sorriso-Valvo, Perrone, Benella, Marco, Bruno, Wang, R{\'e}ville, Baker, Matteini, Yardley, Settino, Sioulas, Alterman, Tenerani, Raines, Holmes, Buchlin, Verdini, Demoulin, Driel-Gesztelyi, Telloni, Consolini, Marcucci, Stangalini, Marino, Fortunato, Mele, Monti, Owen, Louarn, \& Livi}]{damicis_alfvenic_2025}
D{\textquoteright}Amicis, R., Velli, M., Panasenco, O., {et~al.} 2025, Astronomy \& Astrophysics, 693, A243, \dodoi{10.1051/0004-6361/202451686}

\bibitem[{Davis {et~al.}(2023)Davis, Chandran, Bowen, Badman, Wit, Chen, Bale, Huang, Sioulas, \& Velli}]{davis_evolution_2023}
Davis, N., Chandran, B. D.~G., Bowen, T.~A., {et~al.} 2023, The Astrophysical Journal, 950, 154, \dodoi{10.3847/1538-4357/acd177}

\bibitem[{Deng {et~al.}(2023)Deng, Zhou, Dai, Wang, Feng, He, Jiang, Tian, Yang, Hou, Yan, Gan, Bai, Li, Xia, Li, Su, Xiong, Zhang, Zhu, Lin, Zhang, Chen, He, Feng, Zhang, Sun, Zhang, Chen, Tan, Zhang, Yang, \& Wang}]{deng_solar_2023}
Deng, Y., Zhou, G., Dai, S., {et~al.} 2023, 68, \dodoi{10.1360/TB-2022-0674}

\bibitem[{Denskat \& Neubauer(1982)}]{denskat_statistical_1982}
Denskat, K.~U., \& Neubauer, F.~M. 1982, Journal of Geophysical Research: Space Physics, 87, 2215, \dodoi{10.1029/JA087iA04p02215}

\bibitem[{Dmitruk {et~al.}(2014)Dmitruk, Mininni, Pouquet, Servidio, \& Matthaeus}]{dmitruk_magnetic_2014}
Dmitruk, P., Mininni, P.~D., Pouquet, A., Servidio, S., \& Matthaeus, W.~H. 2014, Physical Review E, 90, 043010, \dodoi{10.1103/PhysRevE.90.043010}

\bibitem[{Dorseth {et~al.}(2024)Dorseth, Bourouaine, \& Perez}]{dorseth_1f_2024}
Dorseth, M., Bourouaine, S., \& Perez, J.~C. 2024, The Astrophysical Journal Letters, 974, L34, \dodoi{10.3847/2041-8213/ad81f9}

\bibitem[{Drake {et~al.}(2021)Drake, Agapitov, Swisdak, Badman, Bale, Horbury, Kasper, MacDowall, Mozer, \& Phan}]{drake_switchbacks_2021}
Drake, J.~F., Agapitov, O., Swisdak, M., {et~al.} 2021, Astronomy \& Astrophysics, 650, A2

\bibitem[{Dudok~de Wit {et~al.}(2020)Dudok~de Wit, Krasnoselskikh, Bale, Bonnell, Bowen, Chen, Froment, Goetz, Harvey, Jagarlamudi, Larosa, MacDowall, Malaspina, Matthaeus, Pulupa, Velli, \& Whittlesey}]{dudok_de_wit_switchbacks_2020}
Dudok~de Wit, T., Krasnoselskikh, V.~V., Bale, S.~D., {et~al.} 2020, The Astrophysical Journal Supplement Series, 246, 39, \dodoi{10.3847/1538-4365/ab5853}

\bibitem[{Fox {et~al.}(2016)Fox, Velli, Bale, Decker, Driesman, Howard, Kasper, Kinnison, Kusterer, Lario, Lockwood, McComas, Raouafi, \& Szabo}]{fox_solar_2016}
Fox, N.~J., Velli, M.~C., Bale, S.~D., {et~al.} 2016, Space Science Reviews, 204, 7, \dodoi{10.1007/s11214-015-0211-6}

\bibitem[{Galtier(2016)}]{galtier_introduction_2016}
Galtier, S. 2016, Introduction to {Modern} {Magnetohydrodynamics} (Cambridge: Cambridge University Press), \dodoi{10.1017/CBO9781316665961}

\bibitem[{Goldreich \& Sridhar(1995)}]{goldreich_toward_1995}
Goldreich, P., \& Sridhar, S. 1995, The Astrophysical Journal, 438, 763, \dodoi{10.1086/175121}

\bibitem[{Harris {et~al.}(2020)Harris, Millman, van~der Walt, Gommers, Virtanen, Cournapeau, Wieser, Taylor, Berg, Smith, Kern, Picus, Hoyer, van Kerkwijk, Brett, Haldane, del R{\'i}o, Wiebe, Peterson, G{\'e}rard-Marchant, Sheppard, Reddy, Weckesser, Abbasi, Gohlke, \& Oliphant}]{harris_array_2020}
Harris, C.~R., Millman, K.~J., van~der Walt, S.~J., {et~al.} 2020, Nature, 585, 357, \dodoi{10.1038/s41586-020-2649-2}

\bibitem[{Harten \& Clark(1995)}]{harten_design_1995}
Harten, R., \& Clark, K. 1995, Space Science Reviews, 71, 23, \dodoi{10.1007/BF00751324}

\bibitem[{Heinemann \& Olbert(1980)}]{heinemann_non-wkb_1980}
Heinemann, M., \& Olbert, S. 1980, Journal of Geophysical Research: Space Physics, 85, 1311, \dodoi{10.1029/JA085iA03p01311}

\bibitem[{Hollweg(1978)}]{hollweg_alfven_1978}
Hollweg, J.~V. 1978, Solar Physics, 56, 305, \dodoi{10.1007/BF00152474}

\bibitem[{huang(2025{\natexlab{a}})}]{huang_swturbpy_2025}
huang, z. 2025{\natexlab{a}}, {SWTurbPy},  Zenodo, \dodoi{10.5281/zenodo.16555026}

\bibitem[{huang(2025{\natexlab{b}})}]{huang_hampel_2025}
---. 2025{\natexlab{b}}, Hampel {Filter},  Zenodo, \dodoi{10.5281/zenodo.16555082}

\bibitem[{Huang {et~al.}(2023)Huang, Sioulas, Shi, Velli, Bowen, Davis, Chandran, Matteini, Kang, Shi, Huang, Bale, Kasper, Larson, Livi, Whittlesey, Rahmati, Paulson, Stevens, Case, Wit, Malaspina, Bonnell, Goetz, Harvey, \& MacDowall}]{huang_new_2023}
Huang, Z., Sioulas, N., Shi, C., {et~al.} 2023, The Astrophysical Journal Letters, 950, L8, \dodoi{10.3847/2041-8213/acd7f2}

\bibitem[{Huang {et~al.}(2024)Huang, Velli, Shi, Zhu, Chandran, Bowen, R{\'e}ville, Huang, Hou, Sioulas, Liu, Pulupa, Huang, \& Bale}]{huang_dominance_2024}
Huang, Z., Velli, M., Shi, C., {et~al.} 2024, The Astrophysical Journal Letters, 977, L12, \dodoi{10.3847/2041-8213/ad9271}

\bibitem[{Hudson {et~al.}(2014)Hudson, Svalgaard, \& Hannah}]{hudson_solar_2014}
Hudson, H.~S., Svalgaard, L., \& Hannah, I.~G. 2014, Space Science Reviews, 186, 17, \dodoi{10.1007/s11214-014-0121-z}

\bibitem[{Hunter(2007)}]{hunter_matplotlib_2007}
Hunter, J.~D. 2007, Computing in Science \& Engineering, 9, 90, \dodoi{10.1109/MCSE.2007.55}

\bibitem[{Iroshnikov(1964)}]{iroshnikov_turbulence_1964}
Iroshnikov, P.~S. 1964, {\textbackslash}sovast, 7, 566

\bibitem[{Johnson(1925)}]{johnson_schottky_1925}
Johnson, J.~B. 1925, Physical Review, 26, 71, \dodoi{10.1103/PhysRev.26.71}

\bibitem[{Kasper {et~al.}(2019)Kasper, Bale, Belcher, Berthomier, Case, Chandran, Curtis, Gallagher, Gary, \& Golub}]{kasper_alfvenic_2019}
Kasper, J.~C., Bale, S.~D., Belcher, J.~W., {et~al.} 2019, Nature, 576, 228

\bibitem[{Kasper {et~al.}(2021)Kasper, Klein, Lichko, Huang, Chen, Badman, Bonnell, Whittlesey, Livi, Larson, Pulupa, Rahmati, Stansby, Korreck, Stevens, Case, Bale, Maksimovic, Moncuquet, Goetz, Halekas, Malaspina, Raouafi, Szabo, MacDowall, Velli, Dudok~de Wit, \& Zank}]{kasper_parker_2021}
Kasper, J.~C., Klein, K.~G., Lichko, E., {et~al.} 2021, Physical Review Letters, 127, 255101, \dodoi{10.1103/PhysRevLett.127.255101}

\bibitem[{King \& Papitashvili(2005)}]{king_solar_2005}
King, J.~H., \& Papitashvili, N.~E. 2005, Journal of Geophysical Research: Space Physics, 110, \dodoi{10.1029/2004JA010649}

\bibitem[{Kolmogorov(1941)}]{kolmogorov_local_1941}
Kolmogorov, A. 1941, Akademiia Nauk SSSR Doklady, 30, 301

\bibitem[{Kraichnan(1965)}]{kraichnan_inertialrange_1965}
Kraichnan, R. 1965, \dodoi{10.1063/1.1761412}

\bibitem[{Kruparova {et~al.}(2023)Kruparova, Krupar, Szabo, Pulupa, \& Bale}]{kruparova_quasi-thermal_2023}
Kruparova, O., Krupar, V., Szabo, A., Pulupa, M., \& Bale, S.~D. 2023, The Astrophysical Journal, 957, 13, \dodoi{10.3847/1538-4357/acf572}

\bibitem[{Lam {et~al.}(2015)Lam, Pitrou, \& Seibert}]{lam_numba_2015}
Lam, S.~K., Pitrou, A., \& Seibert, S. 2015, in Proceedings of the {Second} {Workshop} on the {LLVM} {Compiler} {Infrastructure} in {HPC} (Austin Texas: ACM), 1--6, \dodoi{10.1145/2833157.2833162}

\bibitem[{Larosa {et~al.}(2024)Larosa, Chen, McIntyre, Jagarlamudi, \& Sorriso-Valvo}]{larosa_evolution_2024}
Larosa, A., Chen, C. H.~K., McIntyre, J.~R., Jagarlamudi, V.~K., \& Sorriso-Valvo, L. 2024, Astronomy \& Astrophysics, 686, A238, \dodoi{10.1051/0004-6361/202450030}

\bibitem[{Larosa {et~al.}(2021)Larosa, Krasnoselskikh, Wit, Agapitov, Froment, Jagarlamudi, Velli, Bale, Case, Goetz, Harvey, Kasper, Korreck, Larson, MacDowall, Malaspina, Pulupa, Revillet, \& Stevens}]{larosa_switchbacks_2021}
Larosa, A., Krasnoselskikh, V., Wit, T. D.~d., {et~al.} 2021, Astronomy \& Astrophysics, 650, A3, \dodoi{10.1051/0004-6361/202039442}

\bibitem[{Magyar \& Doorsselaere(2022)}]{magyar_phase_2022}
Magyar, N., \& Doorsselaere, T.~V. 2022, The Astrophysical Journal, 938, 98, \dodoi{10.3847/1538-4357/ac8b81}

\bibitem[{Matteini(2019)}]{matteini_rotation_2019}
Matteini, L. 2019, Il Nuovo Cimento C, 42, 1, \dodoi{10.1393/ncc/i2019-19016-y}

\bibitem[{Matteini {et~al.}(2018)Matteini, Stansby, Horbury, \& Chen}]{matteini_1_2018}
Matteini, L., Stansby, D., Horbury, T.~S., \& Chen, C. H.~K. 2018, The Astrophysical Journal, 869, L32, \dodoi{10.3847/2041-8213/aaf573}

\bibitem[{Matteini {et~al.}(2024)Matteini, Tenerani, Landi, Verdini, Velli, Hellinger, Franci, Horbury, Papini, \& Stawarz}]{matteini_alfvenic_2024}
Matteini, L., Tenerani, A., Landi, S., {et~al.} 2024, Physics of Plasmas, 31, 032901, \dodoi{10.1063/5.0177754}

\bibitem[{Matthaeus {et~al.}(2007)Matthaeus, Breech, Dmitruk, Bemporad, Poletto, Velli, \& Romoli}]{matthaeus_density_2007}
Matthaeus, W.~H., Breech, B., Dmitruk, P., {et~al.} 2007, The Astrophysical Journal, 657, L121, \dodoi{10.1086/513075}

\bibitem[{Matthaeus \& Goldstein(1982)}]{matthaeus_stationarity_1982}
Matthaeus, W.~H., \& Goldstein, M.~L. 1982, {\textbackslash}jgr, 87, 10347, \dodoi{10.1029/JA087iA12p10347}

\bibitem[{Matthaeus \& Goldstein(1986)}]{matthaeus_low-frequency_1986}
---. 1986, Physical Review Letters, 57, 495, \dodoi{10.1103/PhysRevLett.57.495}

\bibitem[{Matthaeus {et~al.}(1982)Matthaeus, Goldstein, \& Smith}]{matthaeus_evaluation_1982}
Matthaeus, W.~H., Goldstein, M.~L., \& Smith, C. 1982, Physical Review Letters, 48, 1256, \dodoi{10.1103/PhysRevLett.48.1256}

\bibitem[{Matthaeus \& Lamkin(1986)}]{matthaeus_turbulent_1986}
Matthaeus, W.~H., \& Lamkin, S.~L. 1986, Physics of Fluids, 29, 2513, \dodoi{10.1063/1.866004}

\bibitem[{McComas {et~al.}(2003)McComas, Elliott, Schwadron, Gosling, Skoug, \& Goldstein}]{mccomas_three-dimensional_2003}
McComas, D.~J., Elliott, H.~A., Schwadron, N.~A., {et~al.} 2003, Geophysical Research Letters, 30, \dodoi{10.1029/2003GL017136}

\bibitem[{McIntyre {et~al.}(2023)McIntyre, Chen, \& Larosa}]{mcintyre_properties_2023}
McIntyre, J.~R., Chen, C.~H., \& Larosa, A. 2023, The Astrophysical Journal, 957, 111.
\newblock \url{https://iopscience.iop.org/article/10.3847/1538-4357/acf3dd/meta}

\bibitem[{Milotti(2002)}]{milotti_1f_2002}
Milotti, E. 2002, 1/f noise: a pedagogical review,  arXiv, \dodoi{10.48550/arXiv.physics/0204033}

\bibitem[{Morikawa \& Nakamichi(2023)}]{morikawa_solar_2023}
Morikawa, M., \& Nakamichi, A. 2023, Entropy, 25, 1593, \dodoi{10.3390/e25121593}

\bibitem[{Morton {et~al.}(2025)Morton, Molnar, Cranmer, \& Schad}]{morton_high-frequency_2025}
Morton, R.~J., Molnar, M., Cranmer, S.~R., \& Schad, T.~A. 2025, The Astrophysical Journal, 982, 104, \dodoi{10.3847/1538-4357/adb8df}

\bibitem[{Morton \& Soler(2025)}]{morton_origins_2025}
Morton, R.~J., \& Soler, R. 2025, On the origins of coronal {Alfv{\'e}nic} waves,  arXiv, \dodoi{10.48550/arXiv.2505.08636}

\bibitem[{Morton {et~al.}(2019)Morton, Weberg, \& McLaughlin}]{morton_basal_2019}
Morton, R.~J., Weberg, M.~J., \& McLaughlin, J.~A. 2019, Nature Astronomy, 3, 223, \dodoi{10.1038/s41550-018-0668-9}

\bibitem[{M{\"u}ller {et~al.}(2020)M{\"u}ller, Cyr, Zouganelis, Gilbert, Marsden, Nieves-Chinchilla, Antonucci, Auch{\`e}re, Berghmans, Horbury, Howard, Krucker, Maksimovic, Owen, Rochus, Rodriguez-Pacheco, Romoli, Solanki, Bruno, Carlsson, Fludra, Harra, Hassler, Livi, Louarn, Peter, Sch{\"u}hle, Teriaca, Iniesta, Wimmer-Schweingruber, Marsch, Velli, Groof, Walsh, \& Williams}]{muller_solar_2020}
M{\"u}ller, D., Cyr, O. C.~S., Zouganelis, I., {et~al.} 2020, Astronomy \& Astrophysics, 642, A1, \dodoi{10.1051/0004-6361/202038467}

\bibitem[{Mursula \& Zieger(1996)}]{mursula_135-day_1996}
Mursula, K., \& Zieger, B. 1996, Journal of Geophysical Research: Space Physics, 101, 27077, \dodoi{10.1029/96JA02470}

\bibitem[{Nakagawa \& Levine(1974)}]{nakagawa_dynamics_1974}
Nakagawa, Y., \& Levine, R.~H. 1974, The Astrophysical Journal, 190, 441, \dodoi{10.1086/152896}

\bibitem[{National Academies~of Sciences(2025)}]{national_academies_of_sciences_next_2025}
National Academies~of Sciences, {and}~Medicine, E. 2025, The {Next} {Decade} of {Discovery} in {Solar} and {Space} {Physics}: {Exploring} and {Safeguarding} {Humanity}'s {Home} in {Space} (Washington, DC: The National Academies Press), \dodoi{10.17226/27938}

\bibitem[{Pappalardo {et~al.}(2024)Pappalardo, Buratti, Korth, Senske, Blaney, Blankenship, Burch, Christensen, Kempf, Kivelson, Mazarico, Retherford, Turtle, Westlake, Paczkowski, Ray, Kampmeier, Craft, Howell, Klima, Leonard, Matiella~Novak, Phillips, Daubar, Blacksberg, Brooks, Choukroun, Cochrane, Diniega, Elder, Ernst, Gudipati, Luspay-Kuti, Piqueux, Rymer, Roberts, Steinbr{\"u}gge, Cable, Scully, Castillo-Rogez, Hay, Persaud, Glein, McKinnon, Moore, Raymond, Schroeder, Vance, Wyrick, Zolotov, Hand, Nimmo, McGrath, Spencer, Lunine, Paty, Soderblom, Collins, Schmidt, Rathbun, Shock, Becker, Hayes, Prockter, Weiss, Hibbitts, Moussessian, Brockwell, Hsu, Jia, Gladstone, McEwen, Patterson, McNutt, Evans, Larson, Cangahuala, Havens, Buffington, Bradley, Campagnola, Hardman, Srinivasan, Short, Jedrey, St.~Vaughn, Clark, Vertesi, \& Niebur}]{pappalardo_science_2024}
Pappalardo, R.~T., Buratti, B.~J., Korth, H., {et~al.} 2024, Space Science Reviews, 220, 40, \dodoi{10.1007/s11214-024-01070-5}

\bibitem[{Perez {et~al.}(2021)Perez, Bourouaine, Chen, \& Raouafi}]{perez_applicability_2021}
Perez, J.~C., Bourouaine, S., Chen, C. H.~K., \& Raouafi, N.~E. 2021, Astronomy \& Astrophysics, 650, A22, \dodoi{10.1051/0004-6361/202039879}

\bibitem[{Perrone {et~al.}(2020)Perrone, D{\textquoteright}Amicis, Marco, Matteini, Stansby, Bruno, \& Horbury}]{perrone_highly_2020}
Perrone, D., D{\textquoteright}Amicis, R., Marco, R.~D., {et~al.} 2020, Astronomy \& Astrophysics, 633, A166, \dodoi{10.1051/0004-6361/201937064}

\bibitem[{Pope(2015)}]{pope_turbulent_2015}
Pope, S.~B. 2015, Turbulent flows, 1st edn. (Cambridge: Cambridge Univ. Press)

\bibitem[{Porsche(1981)}]{porsche_helios_1981}
Porsche, H. 1981, 164, 43.
\newblock \url{https://ui.adsabs.harvard.edu/abs/1981ESASP.164...43P}

\bibitem[{Pradata {et~al.}(2025)Pradata, Roy, Matthaeus, Wang, Chhiber, Pecora, \& Yang}]{pradata_observations_2025}
Pradata, R.~A., Roy, S., Matthaeus, W.~H., {et~al.} 2025, The Astrophysical Journal Letters, 984, L23, \dodoi{10.3847/2041-8213/adc9b2}

\bibitem[{Press(1978)}]{press_flicker_1978}
Press, W.~H. 1978, Comments on Astrophysics, 7, 103.
\newblock \url{https://ui.adsabs.harvard.edu/abs/1978ComAp...7..103P}

\bibitem[{Raouafi {et~al.}(2023)Raouafi, Matteini, Squire, Badman, Velli, Klein, Chen, Matthaeus, Szabo, Linton, Allen, Szalay, Bruno, Decker, Akhavan-Tafti, Agapitov, Bale, Bandyopadhyay, Battams, Ber{\v c}i{\v c}, Bourouaine, Bowen, Cattell, Chandran, Chhiber, Cohen, D{\textquoteright}Amicis, Giacalone, Hess, Howard, Horbury, Jagarlamudi, Joyce, Kasper, Kinnison, Laker, Liewer, Malaspina, Mann, McComas, Niembro-Hernandez, Nieves-Chinchilla, Panasenco, Pokorn{\'y}, Pusack, Pulupa, Perez, Riley, Rouillard, Shi, Stenborg, Tenerani, Verniero, Viall, Vourlidas, Wood, Woodham, \& Woolley}]{raouafi_parker_2023}
Raouafi, N.~E., Matteini, L., Squire, J., {et~al.} 2023, Space Science Reviews, 219, 8, \dodoi{10.1007/s11214-023-00952-4}

\bibitem[{Schottky(1918)}]{schottky_uber_1918}
Schottky, W. 1918, Annalen der Physik, 362, 541, \dodoi{10.1002/andp.19183622304}

\bibitem[{Schottky(1922)}]{schottky_zur_1922}
---. 1922, Annalen der Physik, 373, 157, \dodoi{10.1002/andp.19223731007}

\bibitem[{Shaikh(2024)}]{shaikh_turbulence_2024}
Shaikh, Z.~I. 2024, Monthly Notices of the Royal Astronomical Society, 530, 3005, \dodoi{10.1093/mnras/stae897}

\bibitem[{Shi {et~al.}(2020)Shi, Velli, Pucci, Tenerani, \& Innocenti}]{shi_oblique_2020}
Shi, C., Velli, M., Pucci, F., Tenerani, A., \& Innocenti, M.~E. 2020, The Astrophysical Journal, 902, 142, \dodoi{10.3847/1538-4357/abb6fa}

\bibitem[{Shi {et~al.}(2022)Shi, Panasenco, Velli, Tenerani, Verniero, Sioulas, Huang, Brosius, Bale, Klein, Kasper, Wit, Goetz, Harvey, MacDowall, Malaspina, Pulupa, Larson, Livi, Case, \& Stevens}]{shi_patches_2022}
Shi, C., Panasenco, O., Velli, M., {et~al.} 2022, The Astrophysical Journal, 934, 152, \dodoi{10.3847/1538-4357/ac7c11}

\bibitem[{Sioulas {et~al.}(2022)Sioulas, Huang, Velli, Chhiber, Cuesta, Shi, Matthaeus, Bandyopadhyay, Vlahos, Bowen, Qudsi, Bale, Owen, Louarn, Fedorov, Maksimovi{\'c}, Stevens, Case, Kasper, Larson, Pulupa, \& Livi}]{sioulas_magnetic_2022}
Sioulas, N., Huang, Z., Velli, M., {et~al.} 2022, The Astrophysical Journal, 934, 143, \dodoi{10.3847/1538-4357/ac7aa2}

\bibitem[{Sioulas {et~al.}(2023)Sioulas, Velli, Huang, Shi, Bowen, Chandran, Liodis, Davis, Bale, \& Horbury}]{sioulas_evolution_2023}
Sioulas, N., Velli, M., Huang, Z., {et~al.} 2023, The Astrophysical Journal, 951, 141.
\newblock \url{https://iopscience.iop.org/article/10.3847/1538-4357/acc658/meta}

\bibitem[{Taft {et~al.}(1974)Taft, Hickey, Wunsch, \& Baker}]{taft_equatorial_1974}
Taft, B.~A., Hickey, B.~M., Wunsch, C., \& Baker, D.~J. 1974, Deep Sea Research and Oceanographic Abstracts, 21, 403, \dodoi{10.1016/0011-7471(74)90091-6}

\bibitem[{Taylor(1938)}]{taylor_spectrum_1938}
Taylor, G.~I. 1938, Proceedings of the Royal Society of London Series A, 164, 476, \dodoi{10.1098/rspa.1938.0032}

\bibitem[{Tenerani {et~al.}(2021)Tenerani, Sioulas, Matteini, Panasenco, Shi, \& Velli}]{tenerani_evolution_2021}
Tenerani, A., Sioulas, N., Matteini, L., {et~al.} 2021, The Astrophysical Journal Letters, 919, L31, \dodoi{10.3847/2041-8213/ac2606}

\bibitem[{Terres \& Li(2024)}]{terres_investigating_2024}
Terres, M., \& Li, G. 2024, The Astrophysical Journal, 974, 109, \dodoi{10.3847/1538-4357/ad6cd4}

\bibitem[{Tu \& Marsch(1993)}]{tu_model_1993}
Tu, C.-Y., \& Marsch, E. 1993, Journal of Geophysical Research: Space Physics, 98, 1257, \dodoi{10.1029/92JA01947}

\bibitem[{Tu \& Marsch(1995)}]{tu_mhd_1995}
Tu, C.~Y., \& Marsch, E. 1995, Space Science Reviews, 73, 1, \dodoi{10.1007/BF00748891}

\bibitem[{Tu {et~al.}(1989)Tu, Marsch, \& Thieme}]{tu_basic_1989}
Tu, C.-Y., Marsch, E., \& Thieme, K.~M. 1989, Journal of Geophysical Research: Space Physics, 94, 11739, \dodoi{10.1029/JA094iA09p11739}

\bibitem[{Tu {et~al.}(1984)Tu, Pu, \& Wei}]{tu_power_1984}
Tu, C.-Y., Pu, Z.-Y., \& Wei, F.-S. 1984, Journal of Geophysical Research: Space Physics, 89, 9695, \dodoi{10.1029/JA089iA11p09695}

\bibitem[{Velli(1993)}]{velli_propagation_1993}
Velli, M. 1993, Astronomy and Astrophysics, 270, 304.
\newblock \url{https://ui.adsabs.harvard.edu/abs/1993A&A...270..304V}

\bibitem[{Velli {et~al.}(1989)Velli, Grappin, \& Mangeney}]{velli_turbulent_1989}
Velli, M., Grappin, R., \& Mangeney, A. 1989, Physical Review Letters, 63, 1807, \dodoi{chan}

\bibitem[{Velli {et~al.}(1991)Velli, Grappin, \& Mangeney}]{velli_waves_1991}
---. 1991, Geophysical \& Astrophysical Fluid Dynamics, 62, 101, \dodoi{10.1080/03091929108229128}

\bibitem[{Verdini {et~al.}(2012)Verdini, Grappin, Pinto, \& Velli}]{verdini_origin_2012}
Verdini, A., Grappin, R., Pinto, R., \& Velli, M. 2012, The Astrophysical Journal, 750, L33, \dodoi{10.1088/2041-8205/750/2/L33}

\bibitem[{Virtanen {et~al.}(2020)Virtanen, Gommers, Oliphant, Haberland, Reddy, Cournapeau, Burovski, Peterson, Weckesser, Bright, van~der Walt, Brett, Wilson, Millman, Mayorov, Nelson, Jones, Kern, Larson, Carey, Polat, Feng, Moore, VanderPlas, Laxalde, Perktold, Cimrman, Henriksen, Quintero, Harris, Archibald, Ribeiro, Pedregosa, \& van Mulbregt}]{virtanen_scipy_2020}
Virtanen, P., Gommers, R., Oliphant, T.~E., {et~al.} 2020, Nature Methods, 17, 261, \dodoi{10.1038/s41592-019-0686-2}

\bibitem[{Voss \& Clarke(1975)}]{voss_1fnoise_1975}
Voss, R.~F., \& Clarke, J. 1975, Nature, 258, 317, \dodoi{10.1038/258317a0}

\bibitem[{Wang {et~al.}(2024)Wang, Matthaeus, Chhiber, Roy, Pradata, Pecora, \& Yang}]{wang_1f_2024}
Wang, J., Matthaeus, W.~H., Chhiber, R., {et~al.} 2024, \$1/f\$ {Noise} in the {Heliosphere}: {A} {Target} for {PUNCH} {Science},  arXiv, \dodoi{10.48550/arXiv.2409.02255}

\bibitem[{Wang(2024)}]{wang_coronal_2024}
Wang, Y.-M. 2024, Solar Physics, 299, 54, \dodoi{10.1007/s11207-024-02300-3}

\bibitem[{Wenzel {et~al.}(1992)Wenzel, Marsden, Page, \& Smith}]{wenzel_ulysses_1992}
Wenzel, K.~P., Marsden, R.~G., Page, D.~E., \& Smith, E.~J. 1992, Astronomy and Astrophysics Supplement Series, 92, 207.
\newblock \url{https://ui.adsabs.harvard.edu/abs/1992A&AS...92..207W}

\bibitem[{Whang(1973)}]{whang_alfven_1973}
Whang, Y.~C. 1973, Journal of Geophysical Research, 78, 7221, \dodoi{10.1029/JA078i031p07221}

\bibitem[{Wu {et~al.}(2025)Wu, Huang, Yang, He, \& Yuan}]{wu_spectral_2025}
Wu, H., Huang, S., Yang, L., He, J., \& Yuan, Z. 2025, The Astrophysical Journal, 984, 167, \dodoi{10.3847/1538-4357/adcd67}

\bibitem[{Wu {et~al.}(2020)Wu, Tu, Wang, He, \& Yang}]{wu_energy_2020}
Wu, H., Tu, C., Wang, X., He, J., \& Yang, L. 2020, The Astrophysical Journal Letters, 904, L8, \dodoi{10.3847/2041-8213/abc5b6}

\bibitem[{Wu {et~al.}(2021)Wu, Tu, Wang, He, Yang, \& Yao}]{wu_energy_2021}
Wu, H., Tu, C., Wang, X., {et~al.} 2021, The Astrophysical Journal, 912, 84, \dodoi{10.3847/1538-4357/abf099}

\bibitem[{Zank {et~al.}(2020{\natexlab{a}})Zank, Nakanotani, Zhao, Adhikari, \& Kasper}]{zank_origin_2020}
Zank, G.~P., Nakanotani, M., Zhao, L.-L., Adhikari, L., \& Kasper, J. 2020{\natexlab{a}}, The Astrophysical Journal, 903, 1, \dodoi{10.3847/1538-4357/abb828}

\bibitem[{Zank {et~al.}(2020{\natexlab{b}})Zank, Nakanotani, Zhao, Adhikari, \& Telloni}]{zank_spectral_2020}
Zank, G.~P., Nakanotani, M., Zhao, L.-L., Adhikari, L., \& Telloni, D. 2020{\natexlab{b}}, The Astrophysical Journal, 900, 115, \dodoi{10.3847/1538-4357/abad30}

\bibitem[{Zhao {et~al.}(2024)Zhao, Zank, Opher, Zieger, Li, Florinski, Adhikari, Zhu, \& Nakanotani}]{zhao_turbulence_2024}
Zhao, L.-L., Zank, G.~P., Opher, M., {et~al.} 2024, The Astrophysical Journal, 973, 26, \dodoi{10.3847/1538-4357/ad64c8}

\bibitem[{Zhu {et~al.}(2025)Zhu, Zank, Zhao, \& Silwal}]{zhu_radial_2025}
Zhu, X., Zank, G.~P., Zhao, L., \& Silwal, A. 2025, The Astrophysical Journal Letters, 978, L34, \dodoi{10.3847/2041-8213/ada354}

\end{thebibliography}
\bibliographystyle{aasjournal}

\end{CJK*}
\end{document}